\documentclass{article}
\usepackage{amsmath,amssymb,amscd,latexsym,amsthm}
\usepackage{cancel}
\usepackage{hyperref}
\usepackage{epigraph}
\usepackage{appendix}
\usepackage[sorting=none,backend=bibtex]{biblatex}
\usepackage{tikz}
\usepackage{tikz-cd}

\newcommand{\bra}[1]{\langle #1|}
\newcommand{\ket}[1]{|#1\rangle}

\numberwithin{equation}{section}
\author{Peter Woit \\
Department of Mathematics, Columbia University\\
woit@math.columbia.edu}
\title{Euclidean Twistor Unification}
\begin{document}
\maketitle

\nocite{*}

\begin{abstract}

Taking Euclidean signature space-time with its local $Spin(4)=SU(2)_L\times SU(2)_R$  group of space-time symmetries as fundamental, one can consistently gauge the $SU(2)_R$ factor to get a chiral spin connection formulation of general relativity,  and the $SU(2)_L$ factor to get part of the Standard Model gauge fields.  Reconstructing a Lorentz signature theory requires introducing a degree of freedom specifying the imaginary time direction, which will play the role of the Higgs field.

Conformally compactifying $\mathbf R^4$ to $S^4$, one can identify this $S^4$ with the quaternionic projective space $\mathbf {HP}^1$, and the tautological $\mathbf H$ bundle as the bundle of right-handed spinors.  Euclidean twistor geometry is based on the idea that one should work with the projective twistor space $PT=\mathbf {CP}^3$, which is a bundle over $\mathbf {HP}^1$ with fiber $\mathbf {CP}^1$.  A point in the fiber is a complex line $\mathbf C$ inside the $\mathbf H$ specifying the point in $\mathbf {HP}^1$ and defines a complex structure identifying the tangent space $\mathbf R^4$ to $\mathbf {HP}^1$ with $\mathbf C^2$.  The Higgs field specifying the imaginary time direction needs to be taken to be a field on $PT$, lying in this $\mathbf C^2$.  $\mathbf {CP}^3$ comes with an internal $U(1)\times SU(3)$ symmetry at each point, providing the rest of the Standard Model internal symmetries.   The transformation properties of a generation of fermions have a simple expression on $PT$.

This geometry simply encodes  the symmetries and degrees of freedom that go into the Standard Model  and general relativity.  Such a theory is naturally defined on projective twistor space rather than the usual space-time, so will require further development of a gauge theory and spinor field quantization formalism in that context.

\end{abstract}

\section{Introduction}

Quantum field theories in Minkowski space-time suffer from inherent definitional problems, which sometimes can be resolved by defining the theory in terms of an analytic continuation from Euclidean space-time.  The change in space-time signature changes quite a bit the nature of the theory.  Minkowski space-time quantum fields are non-commuting operators satisfying an equation of motion, with free-field Wightman functions that are distributional boundary values of holomorphic functions and no distinguished time direction.  Euclidean space-time quantum fields commute and satisfy no equation of motion, with free-field Schwinger functions that are actual functions and an imaginary time direction specified by the choice of analytic continuation to Minkowski signature.

While in Minkowski space-time one can define the space of states covariantly, starting with a Euclidean theory one needs to pick an imaginary time direction, which gets used to define the state space and a Lorentz-invariant inner product (using Osterwalder-Schrader reflection in the imaginary time direction).     
In the path integral formalism this is the direction perpendicular to the hypersurface used to define states.  If one instead starts with a Minkowski space-time theory and decides to analytically continue to Euclidean space-time, one finds that there is an infinity of possible Euclidean slices of the complexification to use, with each one characterized by the choice of imaginary time direction.  This can be clearly seen in the twistor formalism (see appendix \ref{s:twistors}) where the $SU(2,2)$ conformal symmetry of Minkowski space is determined by a $(2,2)$ signature Hermitian form $\Phi$.   On projective twistor space $PT$ the null space of $\Phi$ is five-dimensional, projecting down to the $\tau=0$ $S^3$ subspace of compactified Euclidean space $S^4$.  Different projections correspond to different choices of an imaginary time parameter $\tau$.

 The Lorentz group $SL(2,\mathbf C)$ is a simple group which acts on the physical state space of the Minkowski space-time theory, while the Euclidean analog $Spin(4)=SU(2)_L\times SU(2)_R$ is a product of two simple groups which does not act on the physical state space.  The quite different nature of these two groups has always made it difficult to understand the relationship between quantum field theories of spinor fields in Minkowski and Euclidean space. For a detailed discussion of these issues, see appendix \ref{s:euclidean-qft}.  Gauging space-time symmetries gives one approach to quantum gravity theories, where again one finds a problematic relationship between Minkowski and Euclidean signature theories.

We will argue here that one should take as fundamental four-dimensional quantum field theory in Euclidean signature, and if one does this, the symmetries and degrees of freedom of the Standard Model and general relativity appear very naturally.   The connection one gets from gauging the $SU(2)_R$ subgroup of $Spin(4)$ can be used to formulate Einstein's equations and general relativity.  The $SU(2)_L$ subgroup plays the role of the internal symmetry of the weak interactions, which after gauging gives part of the Standard Model gauge fields.

Starting with a Euclidean signature theory, definition of the state space and reconstruction of a Minkowski signature theory by analytic continuation require introducing a degree of freedom that breaks the $Spin(4)$ symmetry by picking out an imaginary time direction.  Having such a new degree of freedom also allows one to consistently treat $SU(2)_R$ symmetry as a space-time symmetry, $SU(2)_L$ as an internal symmetry. This degree of freedom has the correct properties to get identified with the Higgs field of the Standard Model. For this to work correctly, one needs to take advantage of another aspect of four-dimensional geometry by using twistor theory.

Penrose's 1967 \cite{penrose-twistors} twistor geometry provides a remarkable alternative to conventional ways of thinking about the geometry of space and time.  In the usual description of space-time as a pseudo-Riemannian manifold, the spinor degree of freedom carried by all matter particles has no simple or natural explanation.   Twistor geometry characterizes a point in Minkowski space-time as a complex 2-plane in $\mathbf C^4$,  with this $\mathbf C^2$ providing tautologically the (Weyl) spinor degree of freedom at the point and an inherent parity-asymmetry.   The $\mathbf C^4$ is the twistor space $T$, and it is often convenient to work with its projective version $PT=\mathbf {CP}^3$, the space of complex lines in  $T$.   Conformal symmetry becomes very simple to understand, with conformal transformations given by linear transformations of $\mathbf C^4$.

Twistor geometry most naturally describes not Minkowski space-time, but its complexification, as the Grassmanian $G_{2,4}(\mathbf C)$ of all complex 2-planes in the twistor space $T$.  This provides a joint complexification of the Euclidean and Minkowski signature spinor and twistor geometries, allowing one to see how they are related by analytic continuation.  Focusing on the Euclidean rather than Minkowski version, it is a remarkable fact that the specific internal symmetry groups and degrees of freedom of the Standard Model appear naturally, unified with the space-time degrees of freedom.  Besides the two $SU(2)$s from $Spin(4)$, since
projective twistor space $PT$ can be thought of as
$$\mathbf {CP}^3=\frac{SU(4)}{U(1)\times SU(3)}$$
there are $U(1)$ and $SU(3)$ internal symmetry groups at each point in projective twistor space.  Lifting the choice of a tangent vector in the imaginary time direction from Euclidean space-time to $PT$, the internal $U(1)\times SU(2)$ acts on this degree of freedom in the same way the Standard Model electroweak symmetry acts on the Higgs field. 

\section{Four dimensional geometry}

\subsection{Four dimensional geometry in terms of two by two matrices}

Instead of describing four-dimensional Euclidean space $\mathbf E^4$ ($\mathbf R^4$ with the usual positive definite norm) by a list $(x_0,x_1,x_2,x_3)$ of four real coordinates, one can work with a subspace $\mathbf E^4\subset M(2,\mathbf C)$ of  the two by two complex matrices, identifying
$$(x_0,x_1,x_2,x_3)\leftrightarrow x=x_0\mathbf 1-i(x_1\sigma_1 +x_2\sigma_2 +x_3\sigma_3)$$
(here $\sigma_j$ are the Pauli matrices).  The norm-squared is then given by
$$|x|^2=\det x$$
Instead of describing the group $SO(4)$ of four-dimensional rotations in terms of orthogonal real four by four matrices, one can now use pairs $(g_L,g_R)$ of $SU(2)$ matrices, acting by
$$x\rightarrow g_Lxg_R^{-1}$$
This action preserves the subspace $\mathbf E^4$, as well as the norm $|x|$.  $(g_L,g_R)$ and $(-g_L,-g_R)$ give the same rotation, and one finds that the product group
$$Spin(4)=SU(2)_L\times SU(2)_R$$ 
is a double cover of the rotation group $SO(4)$.  Dimension four is very special: it is only in this dimension that rotations are not a simple group, but a product of two factors.

Instead of using complex matrices, one can use the algebra $\mathbf H$ of quaternions, identifying
$$(x_0,x_1,x_2,x_3)\leftrightarrow x=x_0\mathbf 1+x_1\mathbf i +x_2\mathbf j +x_3\mathbf k$$
The norm-squared is
$$|x|^2=x\overline x$$
$Spin(4)$ acts as above, except now $SU(2)=Sp(1)$ is the group of unit quaternions, and two such groups $Sp(1)_L$ and $Sp(1)_R$ act indepently by left and right multiplication.  Note that, using either complex matrices or quaternions, the imaginary time direction is distinguished, since it corresponds to the identity matrix.

In special relativity one takes space-time to be not $\mathbf E^4$, but $\mathbf E^{3,1}$, meaning $\mathbf R^4$ with the Minkowski norm. Here again one can take $\mathbf E^{3,1}\subset M(2,\mathbf C)$, identifying
$$(x_0,x_1,x_2,x_3)\leftrightarrow x=-i(x_0\mathbf 1+ x_1\sigma_1 +x_2\sigma_2 +x_3\sigma_3)$$
with the norm-squared again given by the determinant
$$|x|^2=\det x=-x_0^2+x_1^2+x_2^2+x_3^2$$
The spin double cover of the group $SO(3,1)$ of linear transformations preserving the Minkowski norm is $SL(2,\mathbf C)$, with group elements acting by
$$x\rightarrow (g^\dagger)^{-1} xg^{-1}$$

Of less relevance to physics is $\mathbf E^{2,2}\subset M(2,\mathbf C)$, the subspace of real matrices, in which case one identifies
$$(x_0,x_1,x_2,x_3)\leftrightarrow x=\begin{pmatrix}x_0+x_3&x_1+x_2\\x_1-x_2&x_0-x_3\end{pmatrix}$$ 
where
$$|x|^2=\det x=x_0^2-x_1^2+x_2^2-x_3^2$$
The spin double cover of $SO(2,2)$ is
$$Spin(2,2)=SL(2,\mathbf R)_L\times SL(2,\mathbf R)_R$$
given by pairs $g_L,g_R$ of elements of $SL(2,\mathbf R)$ acting by
$$x\rightarrow g_Lxg_R^{-1}$$

If one complexifies (takes complex rather than real linear combinations) the real vector spaces $\mathbf E^4, \mathbf E^{3,1}, \mathbf E^{2,2}$, in each case one gets the same result, the complex vector space $M(2,\mathbf C)$ of all complex two by two matrices.  The group preserving the norm-squared given by the determinant is now $SO(4,\mathbf C)$ which has spin double cover
$$Spin(4,\mathbf C)=SL(2,\mathbf C)_L\times SL(2,\mathbf C)_R$$
given by pairs $g_L,g_R$ of elements of $SL(2,\mathbf C)$ acting by
$$x\rightarrow g_Lxg_R^{-1}$$
($x$ is now an arbitrary complex two by two matrix).

\subsection{Spinors and twistors}

Writing (complexified) four-dimensional vectors as two by two complex matrices identifies vectors as linear maps from one $\mathbf C^2$ (which we'll call $S_R$) to another $\mathbf C^2$ (which we'll call $S_L$).   $S_R$ and $S_L$ are the spinor spaces for four-dimensional geometry, with vectors elements of the space $Hom(S_R,S_L)$ of linear maps.  In the case of $\mathbf E^{2,2}$ one can use $\mathbf R^2$ instead of $\mathbf C^2$, while for $\mathbf E^4$ and $\mathbf E^{3,1}$ one needs $\mathbf C^2$.  The cases $\mathbf E^4$ and $\mathbf E^{3,1}$ are however very different from each other.  For $\mathbf E^4$, $S_L$ and $S_R$ are completely independent spaces, transforming under $Spin(4)$ by two different $SU(2)$ groups.  For $\mathbf E^{3,1}$ on the other hand,  when $g\in Spin(3,1)=SL(2,\mathbf C)$ acts on $S_R$, this determines its action on $S_L$ (by $(g^\dagger)^{-1}$).

In the twistor theory approach to four-dimensional geometry (see Appendix \ref{s:twistors}), points are given by a $\mathbf C^2\subset \mathbf C^4$ ($\mathbf C^4$ is the twistor space $T$), with the $\mathbf C^2$ a spinor space $S_R$ .  Tangent vectors at such a point are linear maps from this $S_R$ to $S_L$ which gets identified with the quotient $T/S_R$.  The four dimensional geometry here is a complex geometry, with $\mathbf E^4, \mathbf E^{3,1}, \mathbf E^{2,2}$ occuring as real four-dimensional subspaces. This provides a context in which the spinor space at a point is tautologically defined, and in which one can study analytic continuation between the Euclidean geometry $\mathbf E^4$ and the Minkowski geometry $\mathbf E^{3,1}$.

The twistor picture naturally includes a much large group of symmetries, the conformal group. One can identify points at infinity in such a way (the conformal compactification) that the complex four-dimensional geometry is  that of $G_{2,4}(\mathbf C)$, the Grassmannian of all $\mathbf C^2\subset \mathbf C^4$.  The group $SL(4,\mathbf C)$ then acts, with subgroups $Spin(5,1), Spin(4,2), Spin(3,3)$ that act as conformal transformations on the compactifications of $\mathbf E^4$ ($S^4$), $\mathbf E^{3,1}$ ($S^3\times S^1$) and $\mathbf E^{2,2}$ ($G_{2,4}(\mathbf R)$) respectively.

\section{Four dimensional Euclidean quantum field theory and gravi-weak unification}

There is a long history of attempts to quantize general relativity in Euclidean space, for a discussion see for instance \cite{gibbons}.   Such an attempt runs into both technical problems and conceptual puzzles.  Here we'll propose a somewhat different context for this problem,  which may shed new light on these issues.

There have been various proposals (see e.g. \cite{alexander-marciano-smolin} and \cite{nesti-percacci}) for unifying the 
weak and gravitational interactions by gauging $SU(2)$ and Lorentz ($SL(2,\mathbf C)$) subgroups of the complexified space-time symmetry group $Spin(4,\mathbf C)$.  We will argue that one should instead work with Euclidean quantum field theory and the symmetry group $Spin(4)=SU(2)_L\times SU(2)_R$.   One can consistently take $SU(2)_L$ to be an internal symmetry, gauge it and construct the usual Yang-Mills $SU(2)$ gauge theory responsible for the weak interactions.  The $SU(2)_R$ will be a space-time symmetry, and gauging it leads to a conventional version of general relativity, expressed in terms of a chiral spin connection.  

The existence of a non-zero distinguished vector $e_0\in Hom(S_R,S_L)$ in the imaginary time direction (necessary for reconstructing a Minkowski space-time theory) allows one to recover the usual geometry of rotations and spin in the spatial directions.  Identifying an arbitary vector $x$ with $e_0^{-1}x\in Hom (S_R,S_R)$ one finds that this transforms under $Spin(4)$ as
$$e_0^{-1}x\rightarrow g_Re_0^{-1}g_L^{-1}g_Lxg_R^{-1}=g_Re_0^{-1}xg_R^{-1}$$
If $x$ is in the $e_0$ direction then $e_0^{-1}x$ is invariant under this action.  If $x$ is a spatial vector, $e_0^{-1}x$ is invariant under $SU(2)_L$, and transforms as a usual $\mathbf E^3$ vector under $SU(2)_R$.  $SU(2)_R$ is thus the spin double over of the $SO(3)$ group of spatial rotations, and $S_R$ is the usual spin representation in three-dimensional space.

\subsection{General relativity in terms of chiral spin connections}

The geometry of a Riemannian manifold $M$ of dimension $n$ can be described in a formalism close to that of gauge theory, by using the principal $SO(n)$ bundle of orthonormal frames (see for example \cite{kobayashi-nomizu}).  On this bundle (or on a spin double-cover) one has two kinds of 1-forms:
\begin{itemize}
\item
Spin connection 1-forms $\omega$ which take values in the Lie algebra $\mathfrak{spin}(n)$.  These describe infinitesimal parallel transport of not just vectors, but also spinors.  These are the usual connection 1-forms one has for any principal $G$-bundle (here $G=SO(n)$ or $Spin(n)$).
\item
Canonical 1-forms $e$ which take values in $\mathbf R^n$, and at a point in the frame bundle give the coordinates of a vector with respect to the orthonormal frame.  These 1-forms are special to frame bundles.
\end{itemize}
In the Palatini formalism for general relativity in four dimensions, one takes as fundamental fields $e,\omega$, with an action of the form
\begin{equation}
\label{eq:palatini}
\int_M \epsilon_{ABCD}e^A\wedge e^B\wedge \Omega^{CD}(\omega)
\end{equation}
where the indices take values $0,1,2,3$ and $\Omega(\omega)$ is the curvature 2-form for the spin connection $\omega$ (like $\omega$, it has values in $\mathfrak {spin}(4)$).   

The equations of motion are then
\begin{itemize}
\item
Varying the $\omega^{AB}$ gives
$$de^A +\omega ^A_B\wedge e^B=0$$
This is the torsion-free condition, determining $\omega$ in terms of $e$ to be the Levi-Civita connection.
\item
Varying the $e^A$ gives the Einstein equations (written in terms of the $e$ instead of the metric).
\end{itemize}

The decomposition $Spin(4)=SU(2)_L\times SU(2)_R$ implies very special properties of the spin connection and Riemannian curvature tensor in four dimensions (for details of the following, see e.g. \cite{atiyah-hitchin-singer}). 
Since
$$\mathfrak{spin}(4)=\mathfrak{su}(2)_R\oplus \mathfrak{su}(2)_L$$
the spin connection and curvature decompose as
$$\omega=\omega_R + \omega_L,\ \ \ \Omega=\Omega_R(\omega_R) +\Omega_L(\omega_L)$$
(the curvature of an $\mathfrak{su}(2)_R$-valued connection is $\mathfrak{su}(2)_R$-valued, similarly for $\mathfrak{su}(2)_L$).  Acting on spinors, $\omega_R$ and $\Omega_R$ take values in $End(S_R)$,  $\omega_L$ and $\Omega_L$ take values in $End(S_L)$

Two-forms on a four-dimensional Riemannian manifold decompose as
$$\Lambda^2(M)=\Lambda^2_+(M)\oplus \Lambda^2_-(M)$$
into $\pm 1$ (self-dual and anti-self-dual) eigenspaces of the Hodge star operator.    Using the identification of tangent vectors with linear maps from $S_R$ to $S_L$ and the identification of tangent vectors and 1-forms given by the metric, one finds that when one writes 2-forms in terms of spinors
$$\Lambda_+^2(M)\subset Hom(S_R,S_R),\ \ \Lambda_-^2(M)\subset Hom(S_L,S_L)$$
Taking this into account, the curvature two-form in four-dimensions has a decomposition
$$\Omega(\omega)=\Omega_{+,R} (\omega)+\Omega_{+,L}(\omega)+\Omega_{-,R}(\omega)+\Omega_{-,L}(\omega)$$

If $\omega$ is torsion-free, then the condition that it be the connection for a solution to the vacuum Einstein equations is
$$\Omega_{-,R}(\omega)=0$$
and this implies that one also has 
$$\Omega_{+,L}(\omega)=0$$
Solutions with $\Omega_{-,L}=0$ will be self-dual (or \lq\lq half-flat").  It's a remarkable aspect of four-dimensional Riemannian geometry that one can express the condition for a connection and curvature to be Einstein just using $(\omega_R, \Omega_R(\omega_R))$.  This leads to the possibility of \lq\lq chiral" formulations of general relativity that use only these degrees of freedom of the geometry.  In particular, one can take as the action that of equation \ref{eq:palatini}, but replacing $\Omega$ by $\Omega_R$.  The equation of motion coming from varying the connection then sets  $\omega_R$ as the $\mathfrak{su}(2)_R$ component of the torsion-free connection.  For an extensive discussion, see \cite{krasnov}.   The ($\omega_L,\Omega_L(\omega_L)$) degrees of freedom make no appearance in such chiral formulations.  The torsion-free condition picks out a specific $\omega_L$, but if one allows torsion then $\omega_L$ can be arbitrary.

In the Hamiltonian rather than covariant formalism, this chiral formulation of gravity is the one studied by Ashtekar and others (for details see for instance \cite{ashtekar1} or \cite{ashtekar2}), giving the starting point for the loop quantum gravity program.  To get from a four-dimensional Lagrangian to a three-dimensional Hamiltonian formalism one needs to specify a time coordinate and a three-dimensional hypersurface $M$.  One can then restrict four-dimensional right-handed spinor fields to $M$, and use the time coordinate to treat them as three-dimensional spinor fields $S$.  The canonical 1-form $\theta$ is also restricted to $M$, giving an $\mathbf R^3$-valued canonical 1-form, where treating tangent vectors as linear maps on spinors, the $\mathbf R^3\subset End(S)$.  

In the Ashtekar variable Hamiltonian formalism the canonical phase space variables in the Euclidean case are the same as for $SU(2)$ Yang-Mills theory.  In the Yang-Mills case these are interpreted as the gauge-field and electric field, in the Ashtekar case as the spin-connection and canonical 1-form.  In both cases there is a Gauss-law constraint, coming from the action of $SU(2)$ gauge transformations.  In the Yang-Mills case the Hamiltonian is the sum of the norm-squares of the electric and magnetic fields.  In the Ashtekar case diffeomorphism invariance gives four extra constraints, three corresponding to translations in space directions, and the fourth to translation in the time direction.  This last constraint sets the Hamiltonian to zero (at least when there are no boundaries).

The use of a chiral formalism does not by itself resolve the well-known problems with quantizing gravity.  The different treatment we propose in the next section for the degrees of freedom of the other chirality and for the imaginary time component of  $e$ may provide some new possibilities to examine.

\subsection{Weak interactions}

In the previous section we have seen that a gravity theory in four dimensions can be expressed in terms of the  $\mathbf R^4$-valued canonical 1-form $\theta$ and the $\mathfrak{su}(2)_R$-valued connection $\omega_R$ for right-handed spinor fields.   The $SU(2)$ part of the Standard Model can be constructed using instead the $\mathfrak{su}(2)_L$-valued connection $\omega_L$ for left-handed spinor fields and one component (the imaginary time component) of the canonical 1-form as Higgs field. These can be given exactly the dynamics of the Standard Model, using the usual Yang-Mills action and the standard Higgs field kinetic term and potential.

There are two problems with this:
\begin{itemize}
\item The electroweak part of the Standard Model has an extra $U(1)$ gauge symmetry, not just a $SU(2)_L$ gauge symmetry.
\item As a 1-form, the Higgs field transforms non-trivially under both $SU(2)_L$ and $SU(2)_R$, but we want it to be invariant under $SU(2)_R$.  The extra $U(1)$ gauge symmetry should combine with $SU(2)_L$ to give a $U(2)$, with the Higgs field transforming as the defining $\mathbf C^2$ representation.
\end{itemize}
A solution to these problems can be found by formulating the theory in twistor space, as described in the next section.

\section{Euclidean twistor theory and the Standard Model}

For a detailed summary of twistor geometry, see section \ref{s:twistors}.  This geometry plays several different roles simultaneously that allow a new sort of unification of fundamental interactions:
\begin{itemize}
\item
It provides a tautological construction of the $\mathbf C^2$ right-handed spinor degree of freedom, since a point in complexified space-time is precisely the  spinor $\mathbf C^2$, realized as a subspace of twistor space $T=\mathbf C^4$.
\item
This complexified space time has Minkowski and Euclidean real slices, allowing an understanding of how analytic continuation relates Minkowski and Euclidean spinor fields.
\item
A point in projective twistor space defines $U(1)$ and $SU(3)$ groups that can be gauged, giving the rest of the internal symmetries of the standard model (besides the $SU(2)_L$ and $SU(2)_R$ of the previous section).
\end{itemize}
We will take Euclidean signature space-time as fundamental, so will be studying just the Euclidean twistors of section \ref{s:euclid} and this will be the geometric arena for unification.  

\subsection{Projective twistors as the bundle of complex structures}

It is well-known that the study of two-dimensional Riemannian geometry acquires very useful new structure if one identifies $\mathbf E^2=\mathbf C$.  To do this one just needs a complex structure: a $J\in SO(2)$ such that $J^2=-1$, which will play the role of multiplication by $i$.  There are two such $J$, of opposite sign.  If one tries to do the same thing in four dimensions, there is an $S^2$ worth of $J\in SO(4)$ satisfying $J^2=-1$, so an $S^2$ worth of inequivalent different ways of giving $\mathbf E^4$ a complex structure.   One way to motivate Euclidean twistors is to argue that one should consider all such complex structures simultaneously.  The projective twistor space $PT$ of the fiber bundle \ref{eq:twistorfibration}
\begin{equation*}
	\begin{tikzcd}
		S^2\arrow[r]& PT=\mathbf {CP}^3 \arrow[d,"\pi"]\\
		& S^4=\mathbf {HP}^1
	\end{tikzcd}
\end{equation*}
is exactly the bundle of all such complex structures for the manifold $S^4$. The fiber above a point $x$ is the $S^2$ of complex structures on the tangent space at $x$. Unlike $S^4$, $PT=\mathbf {CP}^3$ is a complex manifold and can be studied using complex analysis.  More generally, if $M$ is any four-manifold with anti-self-dual metric, its bundle of complex structures (generalizing $PT$), will be a complex manifold.

The fibers of this fibration can also be thought of as the $\mathbf {CP}^1$ of complex lines  $\mathbf C \subset \mathbf C^2=\mathbf H$ in the fiber of the tautological $\mathbf H$ bundle over $\mathbf{HP}^1$.  This $\mathbf C^2$ is the spinor space $S_R$ at the corresponding point, and tangent vectors are linear maps from $S_R$ to $S_L$.   A point in $PT$ is a complex line $l \subset \mathbf C^4$.  This line also lies in $\mathbf C^2=S_R$ and thus gives the point in the fiber $\mathbf {CP}^1$.  Given $l$, we can identify tangent vectors on $S^4$ with 
$$Hom_{\mathbf C}(l,S_L)$$
the complex two dimensional space of  complex linear maps from $l$ to $S_L$.  More explicitly, if one for instance takes as $l$ the complex line generated by $\begin{pmatrix}1\\0\end{pmatrix}$ then one gets the identification
$$(x_0,x_1,x_2,x_3)\in \mathbf R^4 \leftrightarrow \left(x_0\mathbf 1-i(x_1\sigma_1 +x_2\sigma_2 +x_3\sigma_3)\right)\begin{pmatrix}1\\0\end{pmatrix}=
\begin{pmatrix}x_0-ix_3\\-ix_1+x_2\end{pmatrix}$$

Working on $PT$ instead of $S^4$, one can resolve the problems noted with taking $SU(2)_L$ to be the weak interaction gauge symmetry:
\begin{itemize}
\item A point in $PT$ is a complex line $l$, and this defines a $U(1)$ bundle over $PT$, with fiber the unit vectors in $l$. This provides the extra $U(1)$ gauge symmetry needed for an electroweak theory.
\item Tangent vectors to $S^4$ (lifted to $PT$) lie in $Hom_{\mathbf C}(l,S_L)$ and are acted on by the $U(1)$ (through the action on $l$) and $SU(2)_L$ (through the action on $S_L$).  Lifted to $PT$, a vector in the imaginary time direction will transform under $U(1)$ and $SU(2)_L$ as a Higgs field. 
\end{itemize}
As usual in the electroweak theory, the Higgs picks out a specific $U(1)$ subgroup of the $U(2)$ group, the subgroup that leaves the field vector invariant.  The gauge theory of this $U(1)$ group is that of the theory of electromagnetism.

Given a complex structure $J$ acting on $\mathbf R^4$, one can extend its action to the complexification $\mathbf R^4\otimes \mathbf C$ and decompose this space into eigenspaces of $J$ with eigenvalues $+i$ and $-i$ related by complex conjugation
$$\mathbf R^4\otimes\mathbf C=\mathbf C^2 \oplus \overline{\mathbf C^2}$$
The spinor spaces can then be constructed in terms of antisymmetric tensors as
\begin{equation}
\label{eq:spinorsasantisymmetrictensors}
S_R\oplus S_L= \Lambda^*\mathbf C^2 \otimes (\Lambda^2\overline{\mathbf C^2})^{\frac{1}{2}}
\end{equation}
with
$$S_R=(\Lambda^0\mathbf C^2 \oplus \Lambda^2\mathbf C^2)\otimes (\Lambda^2\overline{\mathbf C^2})^{\frac{1}{2}}$$
and
$$S_L=\Lambda^1\mathbf C^2\otimes (\Lambda^2\overline{\mathbf C^2})^{\frac{1}{2}}$$
The $U(2)\subset SO(4)$ subgroup picked out by the choice of $J$ has a double cover $SU(2)\times U(1)\subset Spin(4)$, with the $SU(2)$ identified as $SU(2)_L$ and the $U(1)$ a subgroup of $SU(2)_R$. For much more detail about this construction see chapter 31 of \cite{woit-qmbook}.

At a point on $PT$, one can consider not just the lifted tangent vectors from $S^4$, but also the spinor spaces $S_R$ and $S_L$, now expressed as in equation \ref{eq:spinorsasantisymmetrictensors} using the complex structure given by the point.  We have seen that there is a $U(1)$ group determined by the point on $PT$, and it can be identified with the $U(1)\subset SU(2)_R$ that acts on right-handed spinors.   As representations of $SU(2)_L$ and this $U(1)$, spinors lifted to $PT$ are
$$S_L=\mathbf C^2_0,\ \ S_R=\mathbf C_{-1}\oplus \mathbf C_{+1}$$
where the subscript is the $U(1)$ weight.

\subsection{PT and $SU(3)$ gauge symmetry}

So far we have just been using aspects of twistor geometry that at a point $l\in PT$ involve the fiber $l\subset \mathbf C^4$ of the tautological line bundle $L$ over $PT$, as well as the fibration \ref{eq:twistorfibration} to $S^4$.  Just as in the case of the Grassmanian $Gr_{2,4}(\mathbf C)$ of section \ref{s:g24}, where one could define not just a tautological bundle $S$, but also a quotient bundle $S^\perp$, over $PT$ one has not just $L$, but also a quotient bundle $L^\perp$. This quotient bundle will have a complex $3$-dimensional fiber at $p$ given by $l^\perp=\mathbf C^4/l$.  One can think of $PT$ as 
$$PT=\frac{U(4)}{U(1)\times U(3)}=\frac{SU(4)}{S(U(1)\times U(3))}=\frac{SU(4)}{U(3)}$$
where the $U(1)$ factor acts as unitary transformations on the fiber $l$, while the $U(3)$ acts as unitary transformations on the fiber $\mathbf C^4/l$.  The $SU(3)\subset U(3)$ subgroup provides the color gauge group of the Standard Model, with fermion fields taking values in $l^\perp$ giving the quarks.  For the $U(1)$ to lie in $S(U(1)\times U(3))\subset U(4)$, if it acts with weight $1$ on $l$, it will act with weight $-\frac{1}{3}$ on $l^\perp$.

We now have, at each point on $PT$, a principal bundle with fiber the internal symmetries of the Standard Model, $SU(3)\times SU(2)_L\times U(1)$.  A generation of Standard Model matter fields has exactly the transformation properties under this group of the space of linear maps from $\mathbf C^4$ to itself,  thought of as
$$Hom( l\oplus l^\perp, S_R\oplus S_L)=(l\oplus l^\perp)^*\otimes ( S_L\oplus S_R)$$
One could write this space as
$$(\mathbf C_{-1}\otimes \mathbf C^3_{\frac{1}{3}})\otimes (\mathbf C_0^2 \oplus \mathbf C_{-1}\oplus \mathbf C_{+1})$$
which is 
$$ \mathbf C^2_{-1}\oplus \mathbf C_{-2}\oplus \mathbf C_{0} \oplus (\mathbf C^3\otimes \mathbf C^2)_{\frac{1}{3}} +\mathbf C^3_{-\frac{2}{3}} +\mathbf C^3_{\frac{4}{3}}$$
Here the subscripts are $U(1)$ weights (weak hypercharge), the $\mathbf C^2$ are the fundamental representation of $SU(2)_L$ and the $\mathbf C^3$ are the fundamental representation of $SU(3)$.  For the first generation, the terms above correspond respectively to the fundamental particles 
$$\begin{pmatrix}\nu_e\\ e \end{pmatrix}_L,  e_R, (\nu_e)_R, \begin{pmatrix}u\\ d\end{pmatrix}_L, u_R, d_R$$

Note that the sort of unification of internal symmetries taking place here is quite different than the usual one of GUT theories: there is no attempt to fit all symmetry groups into a single large Lie group. The degrees of freedom of a single generation are unified in quite different way than in the usual  GUT theories (e.g. $SO(10)$ GUT theories, where a generation fits into a spinor representation).

\subsection{Relating theories on $PT$ and space-time}

The unification proposal being made here involves certain changes in the nature of fundamental degrees of freedom with respect to the conventional description of the Standard Model and general relativity. In particular, the $U(1)$ and $SU(3)$ principal bundles on which gauge theory connections are to be defined live on the projective twistor space $PT$, not Euclidean space-time $S^4$.     A full definition of a theory on $PT$ will not be given here and remains to be done, but several indications of how this might work are as follows.

Considering first  free massless matter fields, without coupling to gauge fields or the Higgs field,  note that the single-particle state space $\mathcal H_1$ for a spinor field can be identified with the initial data at $t=0$ for a solution to the Weyl equation. This has the disadvantage of obscuring the Poincar\'e group action on $\mathcal H_1$,  but  in the twistor formalism one can identify this $t=0$ subspace of compactified Minkowski space (an $S^3$) with an equator in compactified Euclidean space $S^4$ that divides the space into upper ($\tau>0$) and lower ($\tau<0$) hemispheres $S^4_+$ and $S^4_-$.
In the coordinates for $S^4$ of equation \ref{eq:quaternioncoordinate}, setting $\tau=x_0=0$ corresponds to the condition that the real part of the numerator vanish, so
$$s_1^\perp\overline{s_1} +s_2^\perp \overline{s_2}+ \overline{s_1^\perp}s_1 +\overline{s_2^\perp}s_2=0$$
Note that (by equation \ref{eq:minkowskiphi}), this is exactly the condition
$$\Phi(s,s)=0$$
that describes the five-dimensional subspace $N=PT_0$ of $PT$ which contains the complex lines corresponding to Minkowski space.  One has the fibration
\begin{equation*}
	\begin{tikzcd}
		\mathbf {CP}^1 \arrow[r]&N=PT_0\arrow[r,hook]\arrow[d]& PT=\mathbf {CP}^3 \arrow[d,"\pi"]\\
		&S^3\arrow[r,hook] & S^4=\mathbf {HP}^1
	\end{tikzcd}
\end{equation*}
as well as
\begin{equation*}
	\begin{tikzcd}
		\mathbf {CP}^1 \arrow[r]&PT_{\pm}\arrow[r]\arrow[d]& PT=\mathbf {CP}^3 \arrow[d,"\pi"]\\
		&S^4_{\pm}\arrow[r] & S^4=\mathbf {HP}^1
	\end{tikzcd}
\end{equation*}

One can study the analytic continuation of solutions of the massless Dirac equation between $S^4_+$ and  compactified Minkowski space (where they are hyperfuctions), by using the Euclidean and Minkowski Penrose transforms to relate both to holomorphic objects on $PT$.  The single-particle state space then will be given by  holomorphic cohomology classes on $PT_+$, the part of $PT$ that projects to the upper hemisphere of $S^4$. The relevant case of the Penrose transform (see section \ref{s:penrose-ward}) is an identification of  the cohomology group $H^1(PT_+,\mathcal O(-3))$ with solutions of the Weyl equation on $S^4_+$.  To get the value of a solution at a point from the cohomology class, one restricts the class to the fiber $\mathbf{CP}^1$ and uses the isomorphism
$$H^1(\mathbf {CP}^1,\mathcal O(-3))=\mathbf C^2$$
For an alternate point of view on this,  \cite{dubois-violette} has a discussion of the relation of the Dirac operator on a manifold such as $S^4$ to the Dolbeault operator on the the projective twistor space.

Introducing gauge fields on $S^4_+$, for anti-self-dual connections on a vector bundle on $S^4_+$, the Penrose-Ward correspondence gives an identification with a holomorphic bundle $F$ on $PT_+$, and solutions of the Weyl equation coupled to that connection are given by cohomology classes $H^1(PT_+,\mathcal O(F(-3)))$.  This however only applies to gauge fields satisfying the anti-self-duality equations rather than the more general Yang-Mills equations.  The idea of studying quantum Yang-Mills theory on twistor space has attracted attention over the years, going back for instance to work of Nair \cite{nair} in 1988.  This works best in a formalism based on expanding about the anti-self-dual theory, as studied by Chalmers-Siegel \cite{chalmers-siegel}. In 2003 Witten \cite{witten-twistor} made major advances in calculating Yang-Mills amplitudes using twistor space, and this led to an active ongoing program of studying such amplitudes that exploits twistor space ideas.  For more of the literature relating supersymmetric Yang-Mills theories and quantum field theories on projective twistor space, see for instance \cite{boels-mason-skinner}, \cite{costello} and the review article \cite{adamo-ym}.  

In the usual formalism, states for the full theory will be functionals of fields (spinor fields, gauge fields, vierbeins, Higgs field) on the $t=0$ subspace of Minkowski space-time.  In a Euclidean twistor formalism,  fields will live on the five-dimensional space $PT_0$ fibered over the $\tau=0$ $S^3$ subspace of Euclidean space-time, with fiber $\mathbf {CP}^1$. One needs to somehow exploit the holomorphic structure of $\mathbf {CP}^1$ to relate fields on $PT_0$ and on $S^3$.
The usual Penrose-Ward correspondence relates bundles on space-time and their pull-backs to projective twistor space $PT$.  The unification proposal here inherently involves  $U(1)$ and $SU(3)$ principal bundles which live on $PT$ and are not pull-backs from space-time, but vary along the $\mathbf {CP}^1$ fiber.

\section{Open problems and speculative remarks}

The unification proposal discussed here is still missing some crucial aspects. Most critically, it is unclear what the origin of generations might be.  This issue is crucial for any hope of understanding where fermion masses and mixing angles come from.  It is possible that the fundamental theory involves not just the usual twistor geometry of $PT$, but should be formulated on the seven-sphere $S^7$, which is a circle bundle over $PT$.  $S^7$ is a remarkably unusual geometric structure, exhibiting a wide range of different symmetry groups, since one has
$$S^7=Spin(8)/Spin(7)=Spin(7)/G_2=Spin(6)/SU(3)=Spin(5)/Sp(1)$$
as well as algebraic structures arising from identifying $S^7$ with the unit octonions.  Our discussion has exploited the last two geometries on $S^7$, not the first two.  The chiral formulation of general relativity may have an interesting formulation using the geometry of $S^7$, see \cite{krasnov} and \cite{herfray}.

The construction of a generation of fermion fields using four-by-four complex matrices with the geometrical interpretation of $Hom( l\oplus l^\perp, S_R\oplus S_L)$ may perhaps be understood as a lift from $S^4$ of the Clifford algebra bundle on $S^4$, which also is a bundle of four-by-four complex matrices.  It is well-known that if one attempts to formulate a lattice gauge theory version of fermions in four Euclidean dimensions, one has problems avoiding multiple copies of spinor representations.  The simplest constuction uses Kogut-Susskind fermions, which can be thought of as taking values in the exterior algebra of differential form (this is the associated graded algebra for the Clifford algebra).  In the usual Kogut-Susskind formalism one is getting multiple copies of the space-time spinors of both chiralities.  The lift of this to $PT$ here would be something different, corresponding to 16 copies of a Weyl spinor, with the Weyl spin degree of freedom coming from the fiber $\mathbf CP^1$ at a point, not the Clifford or exterior algebra.

For another point of view on spaces of solutions of massless field equations and the geometry of $PT$, note that this is an example of the geometric quantization construction of representations of the group $SU(2,2)$.  $PT_+$ is an orbit of $SU(2,2)$ acting on $PT$, providing an important example of the case of \lq\lq minimal" orbits, for which geometric quantization runs into difficult technical problems due to the lack of an appropriate invariant polarization.   For a 1982 history of work on this specific case, see appendix A of \cite{rawnsley}.   It may be that the analytic continuation to the Euclidean space-time perspective will give new insight into these problems.   

The anti-self-duality equations can be formulated as the vanishing of a moment map, for an old speculative discussion of the significance of the path integral over gauge fields in this context see \cite{woit-gaugesymmetry}.  $N=2$ and $N=4$ super Yang-Mills give topological quantum field theories (see \cite{witten-tqft}), with the feature that a \lq\lq twisting" of the space-time symmetry into an internal symmetry plays a crucial role.   One might speculate that these theories have some relation to the hypothetical theory on $PT$ described here.  Twisted $N=4$ super Yang-Mills theories have remarkable properties, including providing a quantum field theory version of geometric Langlands \cite{kapustin-witten},  with  the geometric Langlands program in recent years having given evidence for a dramatic unified perspective relating number theory, geometry and representation theory (and, potentially, physics).

\section{Conclusions}

The main conclusion of this work is that twistor geometry provides a compelling picture of fundamental physics, integrating internal and space-time symmetries, as long as one treats together its Euclidean and Minkowski aspects, related through the projective twistor space PT.   The Euclidean aspect  is crucial for understanding the origin of the Standard Model internal symmetries and the breaking of electroweak symmetry, which is inherent in the Euclidean space-time definition of physical states. 

This picture has many attractive aspects:
\begin{itemize}
\item Spinors are tautological objects (a point in space-time is a space of Weyl spinors), rather than complicated objects that must be separately introduced in the usual geometrical formalism.
\item Analytic continuation between Minkowski and Euclidean space-time can be naturally performed, since twistor geometry provides their joint complexification.
\item Exactly the internal symmetries of the Standard Model occur.
\item Electroweak symmetry breaking has a novel origin in the breaking of Euclidean $SO(4)$ symmetry necessary to define a physical theory.
\item The intricate transformation properties of a generation of Standard Model fermions correspond to a simple construction.
\item One gets a new chiral formulation of gravity, unified with the Standard Model.
\item Conformal symmetry is built into the picture in a fundamental way.
\end{itemize}

Much work remains to be done to explicitly construct and understand a full theory defined on $PT$ that would correspond to the Standard Model and general relativity, with the expected three generations of matter fields. Such a theory might allow understanding of currently unexplained features of the Standard Model, as well as possibly making testable predictions that differ from those of the Standard Model. In particular, the framework proposed is fundamentally chiral as a theory of gravity, not just in the electroweak sector, and this may have observable implications.

\appendix

\appendixpage

\section{Twistor geometry}
\label{s:twistors}

Twistor geometry is a 1967 proposal \cite{penrose-twistors} due to Roger Penrose for a very different way of formulating four-dimensional space-time geometry.  For a detailed expository treatment of the subject, see \cite{ward-wells} (for a version aimed at physicists and applications in amplitude calculations, see \cite{adamo}).   Fundamental to twistor geometry is the twistor space $T=\mathbf C^4$, as well as its projective version, the space $PT=\mathbf {CP}^3$ of complex lines in $T$.  

\subsection{Compactified and complexified space-time}
\label{s:g24}

The relation of twistor space to conventional space-time is that complexified and compactified space-time is identified with the Grassmanian $M=G_{2,4}(\mathbf C)$ of complex two-dimensional linear subspaces in $T$.  A space-time point is thus a $\mathbf C^2$ in $\mathbf C^4$ which tautologically provides the spinor degree of freedom at that point.  The spinor bundle $S$ is the tautological two-dimensional complex vector bundle over $M$ whose fiber $S_m$ at a point $m\in M$  is the $\mathbf C^2$ that defines the point.

The group $SL(4,\mathbf C)$ acts on $T$ and transitively on the spaces $PT$ and $M$ of its complex subspaces.   Points in the Grassmanian $M$ can be represented as elements 
$$\omega =(v_1\otimes v_2- v_2\otimes v_1)\in\Lambda^2(\mathbf C^4)$$
by taking two vectors $v_1,v_2$ spanning the subspace.   $\Lambda^2(\mathbf C^4)$ is six conplex dimensional and scalar multiples of $\omega$ gives the same point in $M$, so $\omega$ identifies $M$ with a subspace of $P(\Lambda^2(\mathbf C^4))=\mathbf {CP}^5$.   Such $\omega$  satisfy the equation
\begin{equation}
	\label{eq:klein}
	\omega\wedge \omega=0
\end{equation}
which identifies (the \lq\lq Klein correspondence") $M$ with a submanifold of $\mathbf {CP}^5$ given by a non-degenerate quadratic form.  Twistors are spinors in six dimensions, with the action of $SL(4,\mathbf C)$
on $\Lambda^2(\mathbf C^4)=\mathbf C^6$ preserving the quadratic form \ref{eq:klein}, and giving the spin double cover homomorphism
$$SL(4,\mathbf C)=Spin(6,\mathbf C)\rightarrow SO(6,\mathbf C)$$

To get the tangent bundle of $M$, one needs not just the spinor bundle $S$, but also another two complex-dimensional vector bundle,  the quotient bundle $S^\perp$ with fiber $S^\perp_m=\mathbf C^4/S_m$.  Then the tangent bundle is
$$TM=Hom(S,S^\perp)=S^*\otimes S^\perp$$
with the tangent space $T_mM$ a four complex dimensional vector space given by $Hom(S_m,S^\perp_m)$, the linear maps from $S_m$ to $S^\perp_m$.

A choice of coordinate chart on $M$ is given by picking a point $m\in M$ and identifying  $S^\perp_m$ with a complex two plane transverse to $S_m$.   The point $m$ will be the origin of our coordinate system, so we will denote $S_m$ by $S_0$ and $S^\perp_m$ by $S^\perp_0$.  Now  $T=S_0\oplus S_0^\perp$  and one can choose basis elements $\mathbf e_1,\mathbf e_2\in S_0,\ \mathbf e_3, \mathbf e_4 \in S_0^\perp$ for $T$.  The coordinate of the two-plane spanned by the columns of 
$$\begin{pmatrix}
	1&0\\
	0&1\\
	z_{01}&z_{01}\\
	z_{10}&z_{11}
\end{pmatrix}$$
will be the $2$ by $2$ complex matrix
$$Z=\begin{pmatrix}
	z_{01}&z_{01}\\
	z_{10}&z_{11}
\end{pmatrix}$$
This coordinate chart does not include all of $M$, since it misses those points in $M$ corresponding to complex two-planes that are not transverse to $S_0^\perp$.   Our interest however will ultimately be not in the global structure of $M$, but in its local structure near the chosen point $m$, which we will study using the  $2$ by $2$ complex matrix $Z$ as coordinates.  When we discuss $M$ we will sometimes not distinguish between $M$ and its local version as a complex four-dimensional vector space with origin of coordinates at $m$.

Writing elements of $T$ as
$$\begin{pmatrix}s_1\\s_2\\s^\perp_1\\s^\perp_2\end{pmatrix}$$
an element of $T$ will be in the complex two plane with coordinate $Z$ when
\begin{equation}
	\label{eq:incidence}
	\begin{pmatrix}s^\perp_1\\s^\perp_2\end{pmatrix}=Z\begin{pmatrix}s_1\\s_2\end{pmatrix}
\end{equation}
This incidence equation characterizes in coordinates the relation between lines (elements of $PT$) and planes (elements of $M$) in twistor space $T$.  We'll sometimes also write this as
$$s^\perp=Zs$$

An $SL(4,\mathbf C)$ determinant $1$ matrix
$$\begin{pmatrix}A&B\\C&D\end{pmatrix}$$
acts on $T$ by
$$\begin{pmatrix} s\\s^\perp\end{pmatrix}\rightarrow \begin{pmatrix} As+ Bs^\perp\\Cs +Ds^\perp\end{pmatrix}$$
On lines in the plane $Z$ this is
$$\begin{bmatrix} s\\Zs\end{bmatrix}\rightarrow \begin{bmatrix} As+ BZs\\Cs +DZs\end{bmatrix}=\begin{bmatrix}(A+BZ)s\\ (C+DZ)(A+BZ)^{-1}(A+BZ)s\end{bmatrix}$$
so the corresponding action on $M$ will be given by
$$ Z\rightarrow (C+DZ)(A+BZ)^{-1}$$

Since $\Lambda^2(S_0)=\Lambda^2(S_0^\perp)=\mathbf C$, $S_0$ and $S_0^\perp$ have (up to scalars) unique choices $\epsilon_{S_0}$ and $\epsilon_{S_0^\perp}$ of non-degenerate antisymmetric bilinear forms, and corresponding choices of  $SL(2,\mathbf C)\subset GL(2,\mathbf C)$ acting on $S_0$ and $S_0^\perp$.  These give (again, up to scalars), a unique choice of a non-degenerate symmetric form on $Hom (S_0,S_0^\perp)$, such that
$$\langle Z,Z\rangle=\det Z$$
The subgroup
$$Spin(4,\mathbf C)=SL(2,\mathbf C)\times SL(2,\mathbf C)\subset SL(4,\mathbf C)$$
of matrices of the form
$$\begin{pmatrix}A&0\\0&D\end{pmatrix}$$
with
$$\det A=\det D=1$$
acts on $M$ in coordinates by
$$Z\rightarrow DZA^{-1}$$
preserving $\langle Z,Z\rangle$.

Besides the spaces $PT$ and $M$ of complex lines and planes in $T$, it is also useful to consider the correspondence space whose elements are complex lines inside a complex plane in $T$.  This space can also be thought of as $P(S)$, the projective spinor bundle over $M$.  There is a diagram of maps 
\begin{equation*}
	\begin{tikzcd}
		&P(S)\arrow[ld,"\mu"']\arrow[rd,"\nu"]\\
		PT&&M
	\end{tikzcd}
\end{equation*}
where $\nu$ is the projection map for the bundle $P(S)$ and $\mu$ is the identification of a complex line in $S$ as a complex line in $T$. $\mu$ and $\nu$ give a correspondence between geometric objects in $PT$ and $M$.  One can easily see that $\mu (\nu^{-1}(m))$ is the complex projective line in $PT$ corresponding to a point $m\in M$  (a complex two plane in $T$ is a complex projective line in $PT$).
In the other direction, $\nu(\mu^{-1})$ takes a point $p$ in $PT$ to $\alpha(p)$, a copy of $\mathbf CP^2$ in $M$, called the \lq\lq $\alpha$-plane" corresponding to $p$.

In our chosen coordinate chart, this diagram of maps is given by
\begin{equation*}
	\begin{tikzcd}
		&\left( Z, s \right)\in P(S)\arrow[ld,"\mu"'] \arrow[rd,"\nu"]\\
		{\begin{bmatrix}s\\ Zs\end{bmatrix}}\in PT & & Z\in M
	\end{tikzcd}
\end{equation*}
The incidence equation \ref{eq:incidence} relating $PT$ and $M$ implies that an $\alpha$-plane is a null plane in the metric discussed above.  Given two points $Z_1,Z_2$ in $M$ corresponding to the same point in $PT$, their difference satisfies
$$s^\perp =(Z_1-Z_2)s=0$$
$Z_1-Z_2$ is not an invertible matrix, so has determinant $0$ and is a null vector.

\subsection {The Penrose-Ward transform}
\label{s:penrose-ward}

The Penrose transform relates solutions of conformally-invariant wave equations on $M$ to sheaf cohomology groups, identifying
\begin{itemize}
	\item
	Solutions to a helicity $\frac{k}{2}$ holomorphic massless wave equation on $U$.
	\item
	The sheaf cohomology group
	$$H^1(\widehat U,\mathcal O(-k-2))$$
\end{itemize}
Here $U\subset M$ and $\widehat U\subset PT$ are open sets related by the twistor correspondence, i.e.
$$\widehat U=\mu (\nu^{-1}(U))$$
We will be interested in cases where $U$ and $\widehat U$ are orbits in $M$ and $PT$ for a real form of $SL(4,\mathbf C)$.  Here $\mathcal O(-k-2)$ is the sheaf of holomorphic sections of the line bundle $L^{\otimes (-k-2)}$ where $L$ is the tautological line bundle over $PT$.  For a detailed discussion, see for instance  chapter 7 of \cite{ward-wells}.

The Penrose-Ward transform is a generalization of the above, introducing a coupling to gauge fields.   One aspect of this is the Ward correspondence, an isomorphism between
\begin{itemize}
	\item
	Holomorphic anti-self-dual $GL(n,\mathbf C)$ connections $A$ on $U\subset M$. 
	\item
	Holomorphic rank $n$ vector bundles $E$ over $\widehat U\subset PT$.
\end{itemize}
Here \lq\lq anti-self-dual" means the curvature of the connection satisfies
$$*F_A=-F_A$$
where $*$ is the Hodge dual. There are some restrictions on the open set $U$, and $E$ needs to be trivial on the complex projective lines corresponding to points $m\in U$.

In one direction, the above isomorphism is due to the fact that the curvature $F_A$ is anti-self-dual exactly when the connection $A$ is integrable on the intersection of an $\alpha$-plane with $U$.  One can then construct the fiber $E_p$ of $E$ at $p$ as the covariantly constant sections of the bundle with connection on the corresponding $\alpha$-plane in $M$.   In the other direction, one can construct a vector bundle $\widetilde E$  on $U$ by taking as fiber at $m\in U$ the holomorphic sections of $E$ on the corresponding complex projective line in $PT$.  Parallel transport in this vector bundle can be defined using the fact that two points $m_1,m_2$ in $U$ on the same $\alpha$-plane correspond to intersecting projective lines in $PT$.   For details, see chapter 8 of \cite{ward-wells} and chapter 10 of \cite{mason-woodhouse}.

Given an anti-self-dual gauge field as above, the Penrose transform can be generalized to a Penrose-Ward transform, relating
\begin{itemize}
	\item
	Solutions to a helicity $k$ holomorphic massless wave equation on $U$, coupled to a vector bundle $\widetilde E$ with anti-self-dual connection $A$.
	\item
	The sheaf cohomology group
	$$H^1(\widehat U, \mathcal O(E)(-k-2))$$
\end{itemize}
For more about this generalization, see \cite{eastwood}.

\subsection{Twistor geometry and real forms}

So far we have only considered complex twistor geometry, in which the relation to space-time geometry is that $M$ is a complexified version of a four real dimensional space-time.  From the point of view of group symmetry, the Lie algebra of $SL(4,\mathbf C)$ is the complexification
$$\mathfrak{sl}(4,\mathbf C)= \mathfrak g \otimes \mathbf C$$
for several different real Lie algebras $\mathfrak g$, which are the real forms of $\mathfrak{sl}(4,\mathbf C)$.  To organize the possibilities, recall that $SL(4,\mathbf C)$ is $Spin(6,\mathbf C)$, the spin group for orthogonal linear transformations in six complex dimensions, so $\mathfrak{sl}(4,\mathbf C)=\mathfrak{so}(6,\mathbf C)$.  If one instead considers orthogonal linear transformations in six real dimensions, there are different possible signatures of the inner product to consider, all of which become equivalent after complexification.  This corresponds to the possible real forms
$$\mathfrak g=\mathfrak{so}(3,3),\mathfrak{so}(4,2),\mathfrak{so}(5,1), \text{and}\  \mathfrak{so}(6)$$
which we will discuss (there's another real form, $\mathfrak{su}(3,1)$, which we won't consider).  For more about real methods in twistor theory, see  \cite{woodhouse}.

\subsubsection{$Spin(3,3)=SL(4,\mathbf R)$}

The simplest way to get a real version of twistor geometry is to take the discussion of section \ref{s:twistors} and replace complex numbers by real numbers.  Equivalently, one can look at subspaces invariant under the usual conjugation, given by the map $\sigma$
$$\sigma\begin{pmatrix}s_1\\s_2\\s^\perp_1\\s^\perp_2\end{pmatrix}=\begin{pmatrix}\overline {s_1}\\\overline {s_2}\\\overline{s^\perp_1}\\ \overline{s^\perp_2}\end{pmatrix}$$
which acts not just on $T$ but on $PT$ and $M$.  The fixed point set of the action on $M$ is $M^{2,2}=G_{2,4}(\mathbf R)$, the Grassmanian of real two-planes in $\mathbf R^4$.  As a manifold, $G_{2,4}(\mathbf R)$  is $S^2\times S^2$, quotiented by a $\mathbf Z_2$.  $M^{2,2}$ is acted on by the group   $Spin(3,3)=SL(4,\mathbf R)$ of conformal transformations.  $\sigma$ acting on $PT$ acts on the $\mathbf CP^1$ corresponding to a  point in $M^{2,2}$ with an action whose fixed points form an equatorial circle.

Coordinates can be chosen as in the complex case, but with everything real.  A point in $M^{2,2}$ is given by a real $2$ by $2$ matrix, which can be written in the form
$$Z=\begin{pmatrix}x_0+x_3&x_1-x_2\\x_1+x_2&x_0-x_3\end{pmatrix}$$
for real numbers $x_0,x_1,x_2,x_3$.  $M^{2,2}$ is acted on by the group   $Spin(3,3)=SL(4,\mathbf R)$ of conformal transformations as in the complex case by
$$ Z\rightarrow (C+DZ)(A+BZ)^{-1}$$
with the subgroup of rotations
$$Z\rightarrow DZA^{-1}$$
for $A,D\in SL(2,\mathbf R)$ given by
$$Spin(2,2)=SL(2,\mathbf R)\times SL(2,\mathbf R)$$
This subgroup preserves
$$\langle Z,Z\rangle= \det Z=x_0^2-x_3^2-x_1^2+x_2^2$$

For the Penrose transform in this case, see Atiyah's account in section 6.5 of \cite{atiyah}.  For the Ward correspondence, see section 10.5 of \cite{mason-woodhouse}.

\subsubsection{$Spin(4,2)=SU(2,2)$}

The real case of twistor geometry most often studied (a good reference is \cite{ward-wells}) is that where the real space-time is the physical Minkowski space of special relativity.   The conformal compactification of Minkowski space is a real submanifold of $M$, denoted here by $M^{3,1}$.  It is acted upon transitively by the conformal group $Spin(4,2)=SU(2,2)$.  This conformal group action on $M^{3,1}$  is most naturally understood using twistor space, as the action on complex planes in $T$ coming from the action of the real form $SU(2,2)\subset SL(4,\mathbf C)$ on $T$.  

$SU(2,2)$  is the subgroup of $SL(4,\mathbf C)$ preserving a real Hermitian form $\Phi$ of signature $(2,2)$ on $T=\mathbf C^4$. In our coordinates for $T$, a standard choice for $\Phi$ is given by
\begin{equation}
	\label{eq:minkowskiphi}
	\Phi \left(\begin{pmatrix}s\\ s^\perp\end{pmatrix},  \begin{pmatrix}s^\prime\\ (s^\perp)^\prime\end{pmatrix}\right)=\begin{pmatrix}\overline s & \overline {s^\perp}\end{pmatrix}\begin{pmatrix}0&1\\1&0\end{pmatrix} \begin{pmatrix}s^\prime\\ (s^\perp)^\prime\end{pmatrix}=s^\dagger (s^\perp)^\prime + (s^\perp)^\dagger s^\prime
\end{equation}
Minkowski space is given by complex planes on which $\Phi=0$, so 
$$\Phi \left(\begin{pmatrix}s\\ Zs\end{pmatrix},  \begin{pmatrix}s\\ Zs\end{pmatrix}\right)=s^\dagger (Z+Z^\dagger)s=0$$
Thus coordinates of points on Minkowski space are anti-Hermitian matrices $Z$, which can be written in the form
$$Z=-i\begin{pmatrix}x_0+ x_3& x_1 -ix_2\\ x_1+ix_2&x_0-x_3\end{pmatrix}=- i (x_0\mathbf 1 + \mathbf x\cdot \mathbf \sigma)$$
where $\sigma_j$ are the Pauli matrices.  The metric is the usual Minkowski metric, since
$$\langle Z,Z\rangle=\det Z = -x_0^2 +x_1^2+x_2^2+x_3^2$$
One can identify compactified Minkowski space $M^{3,1}$ as a manifold with the Lie group $U(2)$ which is diffeomorphic to $(S^3\times S^1)/{\mathbf Z_2}$. The identification of the tangent space with anti-Hermitian matrices reflects the usual identification of the tangent space of $U(2)$ at the identity with the Lie algebra of anti-Hermitian matrices.

$SL(4,\mathbf C)$ matrices are in $SU(2,2)$ when they satisfy
$$\begin{pmatrix}A^\dagger & C^\dagger\\ B^\dagger & D^\dagger\end{pmatrix}\begin{pmatrix}0&1\\ 1&0\end{pmatrix}\begin{pmatrix}A&B\\ C&D \end{pmatrix}=\begin{pmatrix}0&1\\ 1&0\end{pmatrix}$$
The Poincar\'e subgroup $P$ of $SU(2,2)$ is given by elements of $SU(2,2)$ of the form
$$\begin{pmatrix} A&0\\C &(A^\dagger)^{-1}\end{pmatrix}$$
where $A\in SL(2,\mathbf C)$ and $A^\dagger C=- C^\dagger A$.  These act on Minkowski space by
$$Z\rightarrow (C + (A^\dagger)^{-1}Z)A^{-1}= (A^\dagger)^{-1}ZA^{-1}+CA^{-1}$$
One can show that $CA^{-1}$ is anti-Hermitian and gives arbitrary translations on Minkowski space.  The Lorentz subroup is $Spin(3,1)=SL(2,\mathbf C)$ acting by
$$Z\rightarrow (A^\dagger)^{-1}ZA^{-1}$$
Here $SL(2,\mathbf C)$ is acting by the standard representation on $S_0$, and by the conjugate-dual representation on $S_0^\perp$.

Note that, for the action of the Lorentz $SL(2,\mathbf C)$ subgroup, twistors written as elements of $S_0\oplus S_0^{\perp}$ behave like usual Dirac spinors (direct sums of a standard $SL(2,\mathbf C)$ spinor and one in the conjugate-dual representation), with the usual Dirac adjoint, in which the $SL(2,\mathbf C)$-invariant inner product is given by the signature $(2,2)$ Hermitian form
$$\langle \psi_1,\psi_2\rangle=\psi_1^\dagger\gamma_0\psi_2$$
Twistors, with their $SU(2,2)$ conformal group action and incidence relation to space-time points, are however something different than Dirac spinors.

The $SU(2,2)$ action on $M$ has six orbits: $M_{++},M_{--},M_{+0},M_{-0},M_{00}$, where the subscript indicates the signature of $\Phi$ restricted to planes corresponding to points in the orbit.  The last of these is a closed orbit $M^{3,1}$, compactified Minkowski space.    Acting on projective twistor space $PT$, there are three orbits: $PT_+,PT_-, PT_0$, where the subscript indicates the sign of $\Phi$ restricted to the line in $T$ corresponding to a point in the orbit.  The first two are open orbits with six real dimensions, the last a closed orbit with five real dimensions.   The points in compactified Minkowski space $M_{00}=M^{3,1}$ correspond to projective lines in $PT$ that lie in the five dimensional space $PT_0$.  Points in $M_{++}$ and $M_{--}$ correspond to projective lines in $PT_+$ or $PT_-$ respectively.

One can construct infinite dimensional irreducible unitary representations of $SU(2,2)$ using holomorphic geometry on $PT_+$ or $M_{++}$, with the Penrose transform relating the two constructions \cite{epw}. For $\overline{PT_+}$ the closure of the orbit $PT_+$,  the Penrose transform identifies the sheaf cohomology groups $H^1(\overline{PT_+},\mathcal O(-k-2))$ for $k>0$ with holomorphic solutions to the helicity $\frac{k}{2}$ wave equation on $M_{++}$.  Taking boundary values on $M^{3,1}$, these will be real-analytic solutions to the helicity $\frac{k}{2}$ wave equation on compactified Minkowski space.  If one instead considers the sheaf cohomology  $H^1(PT_+,\mathcal O(-k-2))$ for the open orbit  $PT_+$  and takes boundary values on $M^{3,1}$ of solutions on $M_{++}$, the solutions will be hyperfunctions, see \cite{wells-hyperfunctions}. 

The Ward correspondence relates holomorphic vector bundles  on $PT_+$ with anti-self-dual $GL(n,\mathbf C)$ gauge fields on $M_{++}$. However, in this Minkowski signature case, all solutions to the  anti-self-duality equations as boundary values of such gauge fields are complex, so one does not get anti-self-dual gauge fields for compact gauge groups like $SU(n)$.

\subsubsection{$Spin(5,1)=SL(2,\mathbf H)$}
\label{s:euclid}

Changing from Minkowski space-time signature $(3,1)$ to Euclidean space-time signature $(4,0)$, the compactified space-time $M^4=S^4$ is again a real submanifold of $M$.  To understand the conformal group and how twistors work in this case, it is best to work with quaternions instead of complex numbers, identifying $T=\mathbf H^2$.  When working with quaternions, one can often instead use corresponding complex $2$ by $2$ matrices, with a standard choice
$$q=q_0 +q_1\mathbf i +q_2\mathbf j + q_3\mathbf k \leftrightarrow q_0- i(q_1\mathbf \sigma_1 +q_2\mathbf \sigma _2 +q_3\mathbf \sigma_3)$$
For more details of the quaternionic geometry that appears here, see \cite{atiyah} or \cite{salamon}

The relevant conformal group acting on $S^4$ is $Spin(5,1)=SL(2,\mathbf H)$, again best understood in terms of twistors and the linear action of $SL(2,\mathbf H)$ on $T=\mathbf H^2$.  The group $SL(2,\mathbf H)$ is the group of quaternionic $2$ by $2$ matrices satisfying a single condition that one can think of as setting the determinant to one, although the usual determinant does not make sense in the quaternionic case.  Here one can interpret the determinant using the isomorphism with complex matrices, or, at the Lie algebra level, $\frak{sl}(2,\mathbf H)$ is the Lie algebra of $2$ by $2$ quaternionic matrices with purely imaginary trace.

While one can continue to think of points in $S^4\subset M$ as complex two planes, one can also identify these complex two planes as quaternionic lines and $S^4$ as $\mathbf {HP}^1$, the projective space of quaternionic lines in $\mathbf H^2$.  The conventional choice of identification between $\mathbf C^2$ and $\mathbf H$ is
$$s=\begin{pmatrix}s_1\\ s_2\end{pmatrix}\leftrightarrow s=s_1+s_2\mathbf j$$
One can then think of the quaternionic structure as providing an alternate notion of conjugation than the usual one,  given instead by left multiplying by $j\in\mathbf H$.  Using $jzj=-\overline z$ one can show that
\begin{equation}
	\label{eq:conjugation}
	\sigma\begin{pmatrix}s_1\\s_2\\s^\perp_1\\s^\perp_2\end{pmatrix}=\begin{pmatrix}-\overline {s_2}\\\overline {s_1}\\ -\overline{s^\perp_2}\\ \overline{s^\perp_1}\end{pmatrix}
\end{equation}
$\sigma$ satisfies $\sigma^2=-1$ on $T$, so $\sigma^2=1$ on $PT$.  We will see later that while $\sigma$ has no fixed points on $PT$, it does fix complex projective lines.

The same coordinates used in the complex case can be used here,  where now $S_0^\perp$ is a quaternionic line transverse to $S_0$, so coordinates on $T$ are the pair of quaternions  
$$\begin{pmatrix}s\\ s^\perp\end{pmatrix}$$
These are also homogeneous coordinates for points on $S^4=\mathbf {HP}^1$ and our choice of $Z\in \mathbf H$ given by
$$\begin{pmatrix}s\\ Zs \end{pmatrix}$$
as the coordinate in a coordinate system with origin the point with homogeneous coordinates
$$\begin{pmatrix}1\\ 0 \end{pmatrix}$$
The point at $\infty$ will be the one with homogeneous coordinates
$$\begin{pmatrix}0\\  1\end{pmatrix}$$
This is the quaternionic version of the usual sort of choice of coordinates in the case of $S^2=\mathbf CP^1$, replacing complex numbers by quaternions.  The coordinate of a point on $S^4$ with homogeneous coordinates 
$$\begin{pmatrix}s\\ s^\perp\end{pmatrix}$$
will be
\begin{equation}
	\label{eq:quaternioncoordinate}
	s^\perp s^{-1}=\frac{(s_1^\perp +s_2^\perp\mathbf j)(\overline {s_1} -s_1\mathbf j)}{|s_1|^2 +|s_2|^2}
	=\frac{s_1^\perp\overline{s_1} +s_2^\perp \overline{s_2}+(-s_1^\perp s_2+s_2^\perp s_1)\mathbf j}{|s_1|^2 +|s_2|^2}
\end{equation}

A coordinate of a point will now be a quaternion $Z=x_0+x_1\mathbf i +x_2\mathbf j  + x_3\mathbf k$ corresponding to the $2$ by $2$ complex matrix
$$Z= x_0\mathbf 1 -i \mathbf x\cdot \mathbf \sigma=\begin{pmatrix}x_0-ix_3&-ix_1-x_2\\ -ix_1+x_2& x_0+ix_3\end{pmatrix}$$
The metric is the usual Euclidean metric, since
$$\langle Z,Z\rangle=\det Z = x_0^2 +x_1^2+x_2^2+x_3^2$$

The conformal group $SL(2,\mathbf H)$ acts on $T=\mathbf H^2$ by the matrix
$$\begin{pmatrix} A&B\\C&D\end{pmatrix}$$
where $A,B,C,D$ are now quaternions, satisfying together the determinant $1$ condition.
These act on the coordinate $Z$ as in the complex case, by
$$ Z\rightarrow (C+DZ)(A+BZ)^{-1}$$
The Euclidean group in four dimensions will be the subgroup of elements of the form
$$\begin{pmatrix} A&0\\C&D\end{pmatrix}$$
such that $A$ and $D$ are independent unit quaternions, thus in the group $Sp(1)=SU(2)$, and $C$ is an arbitrary quaternion.  The Euclidean group acts by
$$Z\rightarrow DZA^{-1} +CA^{-1}$$
with the spin double cover of the rotational subgroup now $Spin(4)=Sp(1)\times Sp(1)$.  Note that spinors behave quite differently than in Minkowski space: there are independent unitary $SU(2)$ actions on $S_0$ and $S_0^\perp$ rather than a non-unitary $SL(2,\mathbf C)$ action on $S_0$ that acts at the same time on $S_0^\perp$ by the conjugate transpose representation.

The projective twistor space $PT$ is fibered over $S^4$ by complex projective lines 
\begin{equation}
	\label{eq:twistorfibration}
	\begin{tikzcd}
		\mathbf C P^1 \arrow[r]& PT=\mathbf CP^3 \arrow[d,"\pi"]\\
		& S^4=\mathbf HP^1
	\end{tikzcd}
\end{equation}
The projection map $\pi$ is just the map that takes a complex line in $T$ identified with $\mathbf H^2$ to the corresponding quaternionic line it generates (multiplying elements by arbitrary quaternions).  In this case the conjugation map $\sigma$ of \ref{eq:conjugation} has no fixed points on $PT$, but does fix the complex projective line  fibers and thus the points in $S^4\subset M$.  The action of $\sigma$ on a fiber takes a point on the sphere to the opposite point, so has no fixed points.

Note that the Euclidean case of twistor geometry is quite different and much simpler than the Minkowski one.  The correspondence space $P(S)$ (here the complex lines in the quaternionic line specifying a point in $M^4=S^4$) is just $PT$ itself, and the twistor correspondence between $PT$ and $S^4$ is just the projection $\pi$.  Unlike the Minkowski case where the real form $SU(2,2)$ has a non-trivial orbit structure when acting on $PT$, in the Euclidean case the action of the real form $SL(2,\mathbf H)$ is transitive on $PT$.

In the Euclidean case, the projective twistor space has another interpretation, as the bundle of orientation preserving orthogonal complex structures on $S^4$.   A complex structure on a real vector space $V$ is a linear map $J$ such that $J^2=-1$, providing a way to give $V$ the structure of a complex vector space (multiplication by $i$ is multiplication by $J$).  $J$ is orthogonal if it preserves an inner product on $V$.  While on $\mathbf R^2$ there is just one orientation-preserving orthogonal complex structure, on $\mathbf R^4$ the possibilities can be parametrized by a sphere $S^2$.  The fiber $S^2=\mathbf CP^1$  of \ref{eq:twistorfibration} above a point on $S^4$ can be interpreted as the space of orientation preserving orthogonal complex structures on the four real dimensional tangent space to $S^4$ at that point.

One way of exhibiting these complex structures on $\mathbf R^4$ is to identify $\mathbf R^4=\mathbf H$ and then note that, for any real numbers $x_1,x_2,x_3$ such that $x_1^2+x_2^2+x_3^2=1$, one gets an orthogonal complex structure on $\mathbf R^4$ by taking
$$J=x_1\mathbf i +x_2\mathbf j +x_3\mathbf k$$
Another way to see this is to note that the rotation group $SO(4)$ acts on orthogonal complex structures, with a $U(2)$ subgroup preserving the complex structure, so the space of these is
$SO(4)/U(2)$, which can be identified with $S^2$.

More explicitly, in our choice of coordinates, the projection map is
$$\pi:\begin{bmatrix}s\\ s^\perp=Zs\end{bmatrix}\rightarrow Z=\begin{pmatrix}x_0-ix_3&-ix_1-x_2\\ -ix_1+x_2& x_0+ix_3\end{pmatrix}$$
For any choice of $s$ in the fiber above $Z$, $s^\perp$ associates to the four real coordinates specifying $Z$ an element of $\mathbf C^2$.  For instance, if $s=\begin{pmatrix}1,0\end{pmatrix}$, the identification of $\mathbf R^4$ with $\mathbf C^2$ is
$$\begin{pmatrix}x_0\\x_1\\x_2\\x_3\end{pmatrix}\leftrightarrow \begin{pmatrix}x_0-ix_3\\-ix_1+x_2\end{pmatrix}$$
The complex structure on $\mathbf R^4$ one gets is not changed if $s$ gets multiplied by a complex scalar, so it just depends on the point $[s]$ in the $\mathbf CP^1$ fiber. 

For another point of view on this, one can see that for each point $p\in PT$, the corresponding $\alpha$-plane $\nu (\mu^{-1}(p))$ in $M$ intersects its conjugate $\sigma(\nu (\mu^{-1}(p)))$ in exactly one real point, $\pi(p)\in M^4$.  The corresponding line in $PT$ is the line determined by the two points $p$ and $\sigma(p)$.  At the same time, this $\alpha$-plane
provides an identification of the tangent space to $M^4$ at $\pi(p)$ with a complex two plane, the $\alpha$-plane itself.   The $\mathbf CP^1$ of $\alpha$ -planes corresponding to a point in $S^4$ are the different possible ways of identifying the tangent space at that point with a complex vector space.  The situation in the Minkowski space case is quite different: there if $\mathbf CP^1\subset PT_0$ corresponds to a point $Z\in M^{3,1}$, each point $p$ in that $\mathbf CP^1$ gives an $\alpha$-plane intersecting $M^{3,1}$ in a null line, and the $\mathbf CP^1$ can be identified with the \lq\lq celestial sphere" of null lines through $Z$.

In the Euclidean case , the Penrose transform will identify the sheaf cohomology group $H^1(\pi^{-1}(U),\mathcal O(-k-2))$ for $k>0$ with solutions of helicity $\frac{k}{2}$ linear field equations on an open set $U\subset S^4$.   Unlike in the Minkowski space case, in Euclidean space there are $U(n)$ bundles $\widetilde E$ with connections having non-trivial anti-self-dual curvature.  The Ward correspondence between such connections and holomorphic bundles $E$ on $PT$ for $U=S^4$ has been the object of intensive study, see for example Atiyah's survey \cite{atiyah}.   The Penrose-Ward transform identifies
\begin{itemize}
	\item
	Solutions to a field equation on $U$ for sections $\Gamma(S^k\otimes \widetilde E)$, with covariant derivative given by an  anti-self-dual connection $A$, where $S^k$ is the $k$'th symmetric power of the spinor bundle.
	\item
	The sheaf cohomology group
	$$H^1(\widehat U, \mathcal O(E)(-k-2))$$
	where $\widehat U=\pi^{-1} U$.
\end{itemize}
For the details of the Penrose-Ward transform in this case, see \cite{hitchin}.

\subsubsection{$Spin(6)=SU(4)$}

If one picks a positive definite Hermitian inner product on $T$, this determines a subgroup $SU(4)=Spin(6)$ that acts on $T$, and thus on $PT, M$ and $P(S)$.  One has
$$PT=\frac{SU(4)}{U(3)},\ \ M=\frac{SU(4)}{S(U(2)\times U(2))},\  P(S)=\frac{SU(4)}{S(U(1)\times U(2))}$$
and the $SU(4)$ action is transitive on these three spaces.  There is no four real dimensional orbit in $M$ that could be interpreted as a real space-time that would give $M$ after complexification.

In this case the Borel-Weil-Bott theorem relates sheaf-cohomology groups of equivariant holomorphic vector-bundles on $PT, M$ and $P(S)$, giving them explicitly as certain finite dimensional irreducible representations of $SU(4)$.   For more details of the relation between the Penrose transform and Borel-Weil-Bott, see \cite{baston-eastwood}.  The Borel-Weil-Bott theorem \cite{borel-weil-bott} can be recast in terms of index theory, replacing the use of sheaf-cohomology with the Dirac equation \cite{bott}.  For a more general discussion of the relation of representation theory and the Dirac operator, see 
\cite{huang-pandzic}.

\section{Euclidean quantum fields}
\label{s:euclidean-qft}

\epigraph{But   besides  this,  by  freeing   ourselves  from   the   limitation   of   the   Lorentz   group,   which   has   produced   all  the   well-known   difficulties   of  quantum    field   theory,   one  has   here   a  possibility  —  if  this   is  indeed  necessary  —  of   producing   new    theories.     That    is,   one   has   the   possibility   of   con­structing   new   theories   in   the   Euclidean   space  and   then   translating  them  back  into  the  Lorentz  system  to  see  what  they  imply. }{J. Schwinger, 1958\cite{schwinger1958-2}}

\epigraph{Should  the  Feynman path integral  be  well-defined  only  in  Euclidean  space,  as axiomaticians  would  have  it,  then  there  seems  to  exist  a  very  real  problem when dealing with Weyl fields as in the theory of weak interactions or in its unification with QCD.}{P. Ramond, 1981\cite{ramond}}

\epigraph{A certain sense of mystery surrounds Euclidean fermions.} {A. Jaffe and G. Ritter, 2008\cite{jaffe2008}}

That one chirality of Euclidean space-time rotations appears after analytic continuation to Minkowski space-time as an internal symmetry is the most hard to believe aspect of the proposed framework for a unified theory outlined in this paper.  One reason for the very long time that has passed since an earlier embryonic version of this idea (see \cite{woit-ssqm}) is that the author has always found this hard to believe himself.  While the fact that the quantization of Euclidean spinor fields is not straightforward is well-known, Schwinger's early hope that this might have important physical significance (see above) does not appear to have attracted much attention.  In this section we'll outline the basic issue with Euclidean spinor fields, and argue that common assumptions about analytic continuation of the space-time symmetry do not hold in this case.  This issue becomes apparent in the simplest possible context of free field theory.  There are also well-known problems when one attempts to construct a non-perturbative lattice-regularized theory of chiral spinors coupled to gauge fields.

Since Schwinger's first proposal in 1958\cite{schwinger1958-1}, over the years it has become increasingly clear that the quantum field theories governing our best understanding of fundamental physics have a much simpler behavior if one takes time to be a complex variable, and considers the analytic continuation of the theory to imaginary values of the time parameter.   In imaginary time the invariant notion of distance between different points becomes positive, path integrals often become well-defined rather than formal integrals, field operators commute, and expectation values of field operators are conventional functions rather than the boundary values of holomorphic functions found at real time.

While momentum eigenvalues can be arbitrarily positive or negative, energy eigenvalues go in one direction only, which by convention is that of positive energies.   Having states supported only at non-negative energies implies (by Fourier transformation) that, as a function of complex time, states can be analytically continued in one complex half plane, not the other.  A quantum theory in Euclidean space has a fundamental asymmetry in the direction of imaginary time, corresponding to the fundamental asymmetry in energy eigenvalues.

Quantum field theories can be characterized by their $n$-point Wightman (Minkowski space-time) or Schwinger (Euclidean space-time) functions, with the Wightman functions not actual functions, but boundary values of analytic continuations of the Schwinger functions.  For free field theories these are all determined by the $2$-point functions $W_2$ or $S_2$.  The Wightman function $W_2$ is Poincar\'e-covariant, while the Schwinger function $S_2$ is Euclidean-covariant.

This simple relation between the Minkowski and Euclidean space-time free field theories masks a much more subtle relationship at the level of fields, states and group actions on these. In both cases one can construct fields and a Fock space built out of a single-particle state space carrying a representation of the space-time symmetry group.  For the Minkowski theory, fields are non-commuting operators obeying an equation of motion and the single-particle state space is an irreducible unitary representation of the Poincar\'e group.  

The Euclidean theory is quite different.  Euclidean fields commute and do not obey an equation of motion. The Euclidean single-particle state space is a unitary representation of the Euclidean group, but far from irreducible.  It describes not physical states, but instead all possible trajectories in the space of physical states (parametrized by imaginary time).  The Euclidean state space and the Euclidean fields are not in any sense analytic continuations of the corresponding Minkowski space constructions.  For a general theory encompassing the relation between the Euclidean group and Poincar\'e group representations, see \cite{frohlich-osterwalder-seiler}.

One can recover the physical Minkowski theory from the Euclidean theory, but to do so one must break the Euclidean symmetry by choosing an imaginary time direction.  In the following sections we will outline the relation between the Minkowski and Euclidean theories for the cases of the harmonic oscillator, the free scalar field theory, and the free chiral spinor field theory.

\subsection{The harmonic oscillator}

The two-point Schwinger function for the one-dimensional quantum harmonic oscillator of frequency $\omega$ ($\omega>0$) is
$$S_2(\tau)=\frac{1}{(2\pi)(2\omega)}e^{-\omega|\tau|}$$
with Fourier transform
$$\widetilde{S_2}(s)=\frac{1}{\sqrt {2\pi}}\int_{-\infty}^\infty e^{is\tau}S_2(\tau)d\tau=
\frac{1}{(2\pi)^{3/2}}\frac{1}{s^2+\omega^2}$$
In the complex $z=t+i\tau$ plane,  $S_2$ can be analytically continued to the upper half plane as 
$$\frac{1}{(2\pi)(2\omega)}e^{i\omega z}$$
and to the lower half plane as 
$$\frac{1}{(2\pi)(2\omega)}e^{-i\omega z}$$  
The Wightman functions are the analytic continuations to the $t$ (real $z$) axis, so come in two varieties:
$$W_2^-(t)=\lim_{\epsilon\rightarrow 0^+}\frac{1}{(2\pi)(2\omega)}e^{i\omega (t+i\epsilon)}$$
and
$$W_2^+(t)=\lim_{\epsilon\rightarrow 0^+}\frac{1}{(2\pi)(2\omega)}e^{-i\omega (t-i\epsilon)}$$

The conventional interpretation of $W_2^\pm$ is not as functions, but as distributions, given as the boundary values of holomorphic functions.  Alternatively (see appendix \ref{s:hyperfunctions}), one can interpret $W_2^\pm (t)$ as the lower and upper half-plane holomorphic functions defining a hyperfunction. Like distributions, hyperfunctions can be thought of a elements of a dual space to a space of well-behaved test functions, in this case a space of real analytic functions. The Fourier transform of $W_2$ is then the hyperfunction
$$\widetilde{W_2}(E)=\frac{i}{(2\pi)^{3/2}}\frac{1}{\omega^2-E^2}$$
which is a sum of terms $\widetilde{W}^{\pm}_2(E)$ supported at $\omega>0$ and $-\omega<0$. Note that the convention that $e^{-iEt}$ has positive energy means that Fourier transforms of positive energy functions are holomorphic for $\tau<0$.

The physical state space of the harmonic oscillator is determined by the single-particle state space $\mathcal H_1=\mathbf C$.  $\mathcal H_1$ is the state space for a single quantum, it can be thought of as the space of positive energy solutions to the equation of motion
\begin{equation}
	\label{eq:hoequation}
	(\frac{d^2}{dt^2} +\omega^2)\phi=0
\end{equation}
$\mathcal H_1$ can also be constructed using $W_2^+$, by defining 
\begin{align*}
	(f,g)=&\int_{-\infty}^\infty\int_{-\infty}^\infty\overline{ f(t_2)}W_2^+(t_2-t_1)g(t_1)dt_1dt_2\\
	=&\int_{-\infty}^\infty\int_{-\infty}^\infty \overline{f(t_2)}\frac{e^{-i\omega (t_2-t_1)}}{(2\pi)(2\omega)}g(t_1)dt_1dt_2\\
	=& \frac{1}{2\omega}\overline{\widetilde f(\omega)}\widetilde g(\omega)
\end{align*}
for $f,g$  functions in $\mathcal S(\mathbf R)$ and taking the space of equivalence classes 
$$\mathcal H_1=[f]\in\{f\in \mathcal S(\mathbf R)\}/\{(f,f)=0\}$$
One can identify such equivalence classes as
$$[f]=\frac{1}{\sqrt{2\omega}}\widetilde f(\omega)$$
$\mathcal H_1$ is $\mathbf C$ with standard Hermitian inner product
$$\langle [f],[g]\rangle =\frac{1}{2\omega}\overline{\tilde{f}(\omega)}\tilde g(\omega)$$
Note that it doesn't matter whether one takes real or complex valued functions $f$, in either case one gets the same quotient complex vector space $\mathcal H_1$.

Given $\mathcal H_1$ and the inner product $\langle\cdot,\cdot\rangle$, the full state space $\mathcal H$ is an inner product space given by the Fock space construction, with
$$\mathcal H=S^*(\mathcal H_1)=\bigoplus_{k=0}^\infty S^k(\mathcal H_1)$$
In this case the symmetrized tensor product $S^k(\mathcal H_1)$ of $k$ copies of $\mathcal H_1=\mathbf C$ is just again $\mathbf C$, the states with $k$-quanta.  A creation operator $a^\dagger(f)$ (for $f$ real) acts by symmetrized tensor product with $[f]$ and $a(f)$ is the adjoint operator. One can define an operator
$$\widehat \phi (f) = a(f)+ a^\dagger (f)$$
and then 
$$\bra 0 \widehat\phi(f)\widehat\phi(g)\ket 0=\langle [f],[g]\rangle$$
$\widehat \phi (t)$ should be interpreted as an  operator-valued distribution, writing
$$\widehat{\phi}(f)=\int_{-\infty}^\infty \widehat\phi(t)f(t)dt$$
$\widehat \phi $ satisfies the equation of motion \ref{eq:hoequation}.

One can use the Schwinger function $S_2$ to set up a Euclidean (imaginary time $\tau$) Fock space, taking $\mathcal E_1$ to be the space of real-valued functions in $\mathcal S(\mathbf R)$ with inner product
\begin{align*}
	(f,g)_{\mathcal E_1}=& \int_{-\infty}^\infty\int_{-\infty}^\infty f(\tau_2)S_2(\tau_2-\tau_1)g(\tau_1)d\tau_1d\tau_2\\
	=&\int_{-\infty}^\infty\int_{-\infty}^\infty f(\tau_2)\frac{e^{-\omega |\tau_2-\tau_1|}}{(2\pi)(2\omega)}g(\tau_1)d\tau_1d\tau_2
\end{align*}
The Fock space will be
$$\mathcal E=S^*(\mathcal E_1\otimes \mathbf C)$$
based on the complexification of $\mathcal E_1$, with operators $a_E^\dagger(f),a_E(f),\widehat \phi_E(f), \widehat{\phi}_E(\tau)$ defined for $f\in\mathcal E_1$.  Expectation values of products of fields $\widehat \phi_E(f)$ for such real-valued $f$ can be given a probabilistic interpretation (see for instance \cite{glimm-jaffe}).

Note that the imaginary time state space and operators are of a quite different nature than those for real time. The operators $\widehat \phi_E(\tau)$ do not satisfy an equation of motion, and commute for all $\tau$.  They describe not the annihilation and creation of a single quantum, but an arbitrary path in imaginary time of a configuration-space observable.  The state space is much larger than the real-time state space, with $\mathcal E_1$ infinite dimensional as opposed to $\mathcal H_1=\mathbf C$.  

One way to reconstruct the physical real-time theory from the Euclidean theory is to consider the fixed $\tau$ subspace of $\mathcal E_1\otimes \mathbf C$ of complex functions localized at $\tau_0$. Here $f(\tau)=a\delta(\tau-\tau_0)$ for $a\in \mathbf C$ and one defines a Hermitian inner product on $\mathcal E_1\otimes \mathbf C$ by
$$(f,g)_{\mathcal E_1\otimes \mathbf C}=\int_{-\infty}^\infty\int_{-\infty}^\infty \overline{f(\tau_2)}S_2(\tau_2-\tau_1)g(\tau_1)d\tau_1d\tau_2$$

While elements of $\mathcal E_1$ satisfy no differential equation and have no dynamics, one does have an action of time translations on $\mathcal E_1$, with translation by $\tau_0$ taking $a\delta(\tau)$ to $a\delta(\tau-\tau_0)$. Since the inner product satisfies
$$(a\delta(\tau),b\delta(\tau-\tau_0))=abe^{-\omega|\tau_0|}$$
one sees that one can define a Hamiltonian operator generating imaginary time translations on these states by taking $H$ to be multiplication by $\omega$.  The imaginary time translation operator $e^{-\tau_0\omega}$ can be analytically continued from $\tau_0>0$ to real time $t$ as
$$U(t)=e^{-it\omega}$$

Another way to reconstruct the real-time theory is the Osterwalder-Schrader method, which begins by picking out the subspace $\mathcal E^+_1\subset \mathcal E_1\otimes \mathbf C$ of functions supported on $\tau< 0$.  Defining a time reflection operator on $\mathcal E_1$ by
$$\Theta f (\tau)=f(-\tau)$$
one can define
$$(f,g)_{OS}=(\Theta f,g)_{\mathcal E_1\otimes \mathbf C}$$
The physical $\mathcal H_1$ can then be recovered as
$$\mathcal H_1=\frac{\{f\in \mathcal E_1^+\}}{\{(f,f)_{OS}=0\}}$$

Note that for $f,g\in \mathcal E^+_1$ one has (since $f,g$ are supported for $\tau<0$)
\begin{align*}
	(f,g)_{OS}=&\int_{-\infty}^\infty\int_{-\infty}^\infty \overline{f(-\tau_2)}\frac{e^{-\omega |\tau_2-\tau_1|}}{(2\pi)(2\omega)}g(\tau_1)d\tau_1d\tau_2\\
	=&\int_{-\infty}^\infty\int_{-\infty}^\infty \overline{f(\tau_2)}\frac{e^{\omega (\tau_2+\tau_1)}}{(2\pi)(2\omega)}g(\tau_1)d\tau_1d\tau_2\\
	=&\frac{1}{(2\pi)(2\omega)}\int_{-\infty}^\infty \overline{f(\tau_2)}e^{\omega\tau_2}d\tau_2\int_{-\infty}^\infty g(\tau_1)e^{\omega\tau_1}d\tau_1\\
	=&\frac{1}{2\omega}\overline{\widetilde{f}(-i\omega)}\widetilde{g}(-i\omega)
\end{align*}
This gives a map
$$f\in \mathcal E^+_1\rightarrow [f]\in\mathcal H_1$$ 
similar to that of the real-time case
$$f\rightarrow [f]=\frac{1}{\sqrt{2\omega}}\widetilde f(-i\omega)=\frac{1}{\sqrt{2\omega}\sqrt{2\pi}}\int_{-\infty}^\infty e^{\omega\tau}f(\tau)d\tau$$

\subsection{Relativistic scalar fields}

The theory of a  mass $m$ free real scalar field in $3+1$ dimensions can be treated as a straightforward generalization of the above discussion of the harmonic oscillator, treating time in the same way, spatial dimensions with the usual Fourier transform.  Defining
$$\omega_{\mathbf p}=\sqrt{|\mathbf p|^2+m^2}$$
the Fourier transform of the Schwinger function is
$$\widetilde{S_2}(s,\mathbf p)=\frac{1}{(2\pi)^3}\frac{1}{s^2+\omega_{\mathbf p}^2}$$
and the Schwinger function itself is
\begin{align*}
	S_2(\tau,\mathbf x)=&\frac{1}{(2\pi)^2}\int_{\mathbf R^4}e^{i(\tau s +\mathbf x\cdot\mathbf p)}\frac{1}{(2\pi)^3}\frac{1}{s^2+\omega_{\mathbf p}^2}ds d^3\mathbf p\\
	=&\frac {m}{(2\pi)^3\sqrt{\tau^2+|\mathbf x|^2}}K_1(m\sqrt{\tau^2+|\mathbf x|^2})
\end{align*}
where $K_1$ is a modified Bessel function.  This has an analytic continuation to the $z=t+i\tau$ plane, with branch cuts on the $t$ axis from $|\mathbf x|$ to $\infty$ and $-|\mathbf x|$ to $-\infty$.

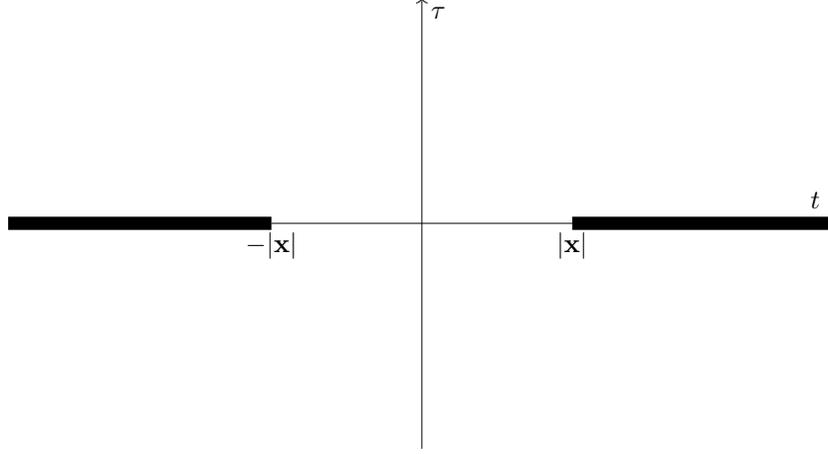
\begin{tikzpicture}
	\draw [very thin] (-2,0)node [align=center, below] {$-|\mathbf x|$} -- (2,0) node [align=center, below] {$|\mathbf x|$};
	\draw [line width=5pt] (-5.5,0) -- (-2,0);
	\draw [line width=5pt] (2,0) -- (5.5,0) node [above left]  {$t$};
	\draw [very thin, ->] (0,-3) -- (0,3) node [below right] {$\tau$};
\end{tikzpicture}
\bigskip

The Wightman function $W_2^+(t,\mathbf x)$ will be defined as the limit of the analytic continuation of $S_2$ as one approaches the $t$-axis from negative values of $\tau$.  This will be analytic for spacelike $t<|\mathbf x|$, but will approach a branch cut for timelike $t>|\mathbf x|$.  The Fourier transform of $W_2^\pm$ will be, as a hyperfunction (in the time-energy coordinate)
$$\widetilde W_2(p)=\frac{1}{(2\pi)^3}\frac{i}{\omega_{\mathbf p}^2-E^2}$$
or, as a distribution, the delta-function distribution 
$$\widetilde W^+_2(p)=\frac{1}{(2\pi)^2}\theta(E)\delta(E^2-\omega_{\mathbf p}^2)$$
on the positive energy mass shell $E=+\mathbf \omega_{\mathbf p}$.  Here $W_2^+(x)$ is
$$W_2^+(t,\mathbf x)=\frac{1}{(2\pi)^4}\int_{\mathbf R^3}\frac{1}{2\omega_{\mathbf p}}e^{-i\omega_{\mathbf p}t}e^{i\mathbf p\cdot\mathbf x}d^3\mathbf p$$

As in the harmonic oscillator case, one can use it to reconstruct the single particle state space $\mathcal H_1$, defining
$$(f,g)=\int_{\mathbf R^4}\int_{\mathbf R^4}f(x)W_2^+(x-y)g(y)d^4xd^4y$$
for $f,g\in \mathcal S(\mathbf R^4)$ ($\mathbf R^4$ is Minkowski space), and equivalence classes
$$\mathcal H_1=[f]\in\{f\in \mathcal S(\mathbf R^4)\}/\{(f,f)=0\}$$

The inner product on $\mathcal H_1$ is given by
$$\langle [f],[g]\rangle =\int_{\mathbf R^4}\theta(E)\delta(E^2-\omega_{\mathbf p}^2)\overline{\widetilde f(p)}\widetilde g(p)d^4p$$
where $p=(E,\mathbf p)$ and $\theta$ is the Heaviside step function.  Elements $[f]$ of $\mathcal H_1$ can be represented by functions $\widetilde f$ on $\mathbf R^3$ of the form
$$\widetilde f(\mathbf p)=\widetilde f(\omega_{\mathbf p},\mathbf p)$$
In this representation, $\mathcal H_1$ has the Lorentz-invariant Hermitian inner product
$$\langle [f],[g]\rangle=\int_{\mathbf R^3}\overline{\widetilde f(\mathbf p)}\widetilde g(\mathbf p)\frac{d^3\mathbf p}{2\omega_{\mathbf p}}$$

Using the Fock space construction (as in the harmonic oscillator case, where $\mathcal H_1=\mathbf C$), the full physical state space is
$$\mathcal H=S^*(\mathcal H_1)=\bigoplus_{k=0}^\infty S^k(\mathcal H_1)$$
with creation operators $a^\dagger(f)$ acting by symmetrized tensor product with $[f]$.  $a(f)$ is the adjoint operator and one can define field operators by
$$\widehat \phi(f)=a(f) +a^\dagger(f)$$
Writing these distributions as $\widehat \phi(t,\mathbf x)$, one recovers the usual description of Wightman functions as 
$$W_2^+(x-y)=\bra{0}\widehat \phi(x)\widehat\phi(y)\ket{0}$$
The operators $\widehat\phi(x)$ satisfy the equation of motion
$$\left(\frac{\partial ^2}{\partial t^2} -\Delta +m^2\right)\widehat\phi=0$$
and $\widehat\phi(x),\widehat\phi(y)$ commute for $x$ and $y$ space-like separated, but not for time-like separations (due to the branch cuts described above).

The Euclidean (imaginary time) theory has the Fock space
$$\mathcal E=S^*(\mathcal E_1\otimes \mathbf C)$$
where $\mathcal E_1$ is  the space of real-valued functions in $\mathcal S(\mathbf R^4)$ (now $\mathbf R^4$ is Euclidean space) with inner product
$$(f,g)_{\mathcal E_1}= \int_{\mathbf R^4}\int_{\mathbf R^4} f(x)S_2(x-y)g(y)d^4xd^4y$$
This Fock space comes with operators $a_E^\dagger(f),a_E(f),\widehat \phi_E(f), \widehat{\phi}_E(x)$ defined for $f\in\mathcal E_1$.  Expectation values of products of fields $\widehat \phi_(f)$ for such real-valued $f$ can be given a probabilistic interpretation in terms of a Gaussian measure on the distribution space $\mathcal S^\prime(\mathbf R^4)$ (for details, see \cite{glimm-jaffe}).

As in the harmonic oscillator case, there are two ways to recover the real time theory from the Euclidean theory.  In the first, one takes $\mathcal H_1\subset \mathcal E_1$ to be the functions on Euclidean space-time localized at a specific value of $\tau$, say $\tau=0$, of the form
$$f(\tau,\mathbf x)=\frac{1}{2\pi}\delta(\tau)F(\mathbf x)$$
Evaluating the inner product for these, one finds
$$(f,g)_{\mathcal E_1}=\int_{\mathbf R^3}\overline{\widetilde F(\mathbf p)}\widetilde{G}(\mathbf p)\frac{d^3\mathbf p}{2\omega_{\mathbf p}}$$
which is the usual Lorentz-invariant inner product.  The rotation group $SO(3)$ of spatial rotations acts on this $\tau=0$ subspace of $\mathcal E_1$ and this action passes to an action on the physical $\mathcal H_1$.  Time translations act on $\mathcal E_1$ and one can use the infinitesimal action of such translations to define the Hamiltonian operator on $\mathcal H_1$.

To recover the physical state space from the Euclidean theory by the Osterwalder-Schrader method, one has to start by picking an imaginary time direction in the Euclidean space $\mathbf R^4$, with coordinate $\tau$.  One can then
restrict to the subspace $\mathcal E^+_1\subset \mathcal E_1$ of functions supported on $\tau<0$.  Defining a time reflection operator on $\mathcal E_1$ by
$$\Theta f (\tau,\mathbf x)=f(-\tau,\mathbf x)$$
one can define
$$(f,g)_{OS}=(\Theta f,g)_{\mathcal E_1}$$
The physical $\mathcal H_1$ can be recovered as
$$\mathcal H_1=\frac{\{f\in \mathcal E_1^+\}}{\{(f,f)_{OS}=0\}}$$

In both the Euclidean and Minkowski space-time formalisms one has a unitary representation of the space-time symmetry groups (the Euclidean group $E(4)$ and the Poincar\'e group $P$ respectively) on the spaces $\mathcal E_1,\mathcal H_1$ and the corresponding Fock spaces.  In the Minkowski space-time case this is an irreducible representation, while in the Euclidean case it is far from irreducible, and the representations in the two cases are not in any sense analytic continuations of each other.

The spatial Euclidean group $E(3)$ is in both $E(4)$ and $P$, and the two methods for passing from the Euclidean to Minkowski space theory preserve this group action.  For translations in the remaining direction, one can fairly readily define the Hamiltonian operator using the semi-group of positive imaginary time translations in Euclidean space, then multiply by $i$ and show that this generates real time translations in Minkowski space-time.

More delicate is the question of what happens for group transformations in other directions in $SO(3,1)$ (the boosts) and $SO(4)$.  In the Minkowski theory, boosts act on $\mathcal H_1$, preserving the inner product, so one has a unitary action of the Poincar\'e group on $\mathcal H_1$ and from this on the full state space (the Fock space).  But while elements of $SO(4)$ not in the spatial $SO(3)$ act on $\mathcal E_1$ preserving $(\cdot,\cdot)_{\mathcal E_1}$, they do not preserve the positive time subspace $\mathcal E_1^+$ and do not commute with the time reflection operator $\Theta$.  One can construct operators on $\mathcal E_1^+$ giving infinitesimal generators corresponding to directions in the Lie algebra complementary to the Lie algebra of $SO(3)$, and then show that these can be analytically continued and exponentiated to give the action of boosts on $\mathcal H_1$. That this can be done was first shown by Klein and Landau in 1982 (by a not completely straight-forward argument, see \cite{klein-landau}).

\subsection{Spinor fields}

While scalar field theories and pure gauge theories have well-understood and straightforward formulations in Euclidean space-time, the question of how to define spinor quantum field theories in Euclidean space-time has always been (see the quote from Jaffe and Ritter above) much more problematic.  At the end of this paper one can find a fairly complete bibliography of attempts to address this question over the years, none of which provide a fully satisfactory answer.   Schwinger's earliest work argued that in Euclidean space a doubling of the spinor degrees of freedom was necessary, and a version of Euclidean spinor fields due to Osterwalder and Schrader \cite{osterwalder1973-2} that includes such a doubling has been the conventionally accepted best solution to the definitional problem.

We'll consider the theory of a chiral (Weyl) spinor field in Minkowski space, and then see what problems arise when one tries to find a corresponding Euclidean field theory.  It is well-known (see the quote at the beginning of this section from \cite{ramond}) that a problem arises immediately if one tries to write down a Euclidean path integral for such a theory: there is no way to write an $SO(4)$ invariant Lagrangian just using one chirality.

The equation of motion for a right-handed Weyl spinor is
$$\left(\frac{\partial}{\partial t}+\boldsymbol{\sigma}\cdot \boldsymbol{\nabla}\right)\psi(t,\mathbf x)=0$$
or, in energy-momentum space
\begin{equation}
	\label{eq:weyl-momentum}
	(E-\boldsymbol{\sigma}\cdot \mathbf p)\widetilde\psi(E,\mathbf p)=0
\end{equation}
Since one has
$$(E+\boldsymbol{\sigma}\cdot \mathbf p)(E-\boldsymbol{\sigma}\cdot \mathbf p)=E^2-|\mathbf p|^2$$
solutions in energy-momentum space will also satisfy
$$(E^2-|\mathbf p|^2)\widetilde\psi(E,\mathbf p)=0$$
and be supported on the light-cone $E=\pm |\mathbf p|$.

The momentum space Wightman function for the Weyl spinor theory will be the hyperfunction
$$\widetilde{W_2}(E,\mathbf p)=\frac{-i}{(2\pi)^3}\frac{1}{E-\boldsymbol{\sigma}\cdot \mathbf p}=\frac{-i}{(2\pi)^3}\frac{E+\boldsymbol{\sigma}\cdot \mathbf p}{E^2-|\mathbf p|^2}$$
or equivalently the distribution
$$\widetilde{W^+_2}(E,\mathbf p)=\frac{1}{(2\pi)^2}\theta(E)(E+\boldsymbol{\sigma}\cdot \mathbf p)\delta (E^2-|\mathbf p|^2)$$
This is matrix-valued, and on solutions to \ref{eq:weyl-momentum} gives the inner product
\begin{align*}
	\langle \widetilde\psi_1,\widetilde \psi_2\rangle=&\int_{\mathbf R^4}\widetilde \psi_1^\dagger(E,\mathbf p)(E+\boldsymbol{\sigma}\cdot \mathbf p)\widetilde \psi_2(E,\mathbf p)\theta(E)\delta (E^2-|\mathbf p|^2)dEd^3\mathbf p\\
	=&\int_{\mathbf R^3}\widetilde \psi_1^\dagger(\mathbf p)(|\mathbf p|+\boldsymbol{\sigma}\cdot \mathbf p)\widetilde \psi_2(\mathbf p)\frac{d^3\mathbf p}{2|\mathbf p|}\\
	=&\int_{\mathbf R^3}\widetilde \psi_1^\dagger(\mathbf p)\widetilde \psi_2(\mathbf p)d^3\mathbf p
\end{align*}
Here $\widetilde\psi(\mathbf p)=\widetilde\psi(|\mathbf p|,\mathbf p)$.

The last expression is manifestly invariant under spatial ($Spin(3)$) rotations, but not Lorentz ($Spin(3,1)=SL(2,\mathbf C)$) transformations.  One can see Lorentz invariance using the  first expression, since for $\Omega\in SL(2,\mathbf C)$ one has
$$(\Omega^\dagger)^{-1}(E+\boldsymbol{\sigma}\cdot \mathbf p)\Omega^{-1}=E^\prime+\boldsymbol{\sigma}\cdot \mathbf p^\prime$$
where $E^\prime, \mathbf p^\prime$ are the Lorentz-transformed energy-momenta
$$(E^\prime,\mathbf p^\prime)=\Lambda^{-1}\cdot (E,\mathbf p)$$
($\Lambda\in SO(3,1)$ corresponds to $\Omega\in Spin(3,1)$ in the spin double cover).

Note that the operator $E+\boldsymbol{\sigma}\cdot \mathbf p$ is just the momentum space identification of Minkowski space-time $\mathbf R^{3,1}$ with $2$ by $2$ hermitian matrices:
$$x=(t,x_1,x_2,x_3)\leftrightarrow M=\begin{pmatrix}t+ x_3& x_1 -ix_2\\ x_1+ix_2&t-x_3\end{pmatrix}$$
with the Minkowski norm given by $-\text{det} M$.  One can identify complexified Minkowski space-time $\mathbf R^{3,1}\otimes \mathbf C=\mathbf C^4$ with all $2$ by $2$ complex matrices by:
$$(t+i\tau,z_1,z_2,z_3)\leftrightarrow M=\begin{pmatrix}t+i\tau+ z_3& z_1 -iz_2\\ z_1+iz_2&t+i\tau-z_3\end{pmatrix}$$
Euclidean space-time $\mathbf R^4$ will get identified with complex matrices of the form
$$(\tau,x_1,x_2,x_3)\leftrightarrow M=\begin{pmatrix}i\tau+ x_3& x_1 -ix_2\\ x_1+ix_2&i\tau-x_3\end{pmatrix}$$
and analytic continuation between Euclidean and Minkowski space takes place on functions of such matrices.

The group $Spin(4,\mathbf C)=SL(2,\mathbf C)\times SL(2,\mathbf C)$ acts on complex matrices by
$$M\rightarrow g_L M g_R^{-1}$$
preserving the determinant (here $g_L,g_R\in SL(2,\mathbf C)$).  The subgroup $SL(2,\mathbf C)$ such that $g_R=(g_L^\dagger)^{-1}$ is the Lorentz group $Spin(3,1)$ that preserves Minkowski space-time, the subspace of hermitian matrices. The subgroup 
$$SU(2)_L\times SU(2)_R=Spin(4)$$
such that $g_L\in SU(2)_L$ and $g_R\in SU(2)_R$ preserves the Euclidean space-time.

If one tries to find a Schwinger function $S_2$ related by analytic continuation to $W_2$ for the Weyl spinor theory, the factor $E+\boldsymbol{\sigma}\cdot \mathbf p$ in the expression for $W_2$ causes two sorts of problems:
\begin{itemize}
	\item After analytic continuation to Euclidean space-time it takes spinors transforming under $SU(2)_R$ to spinors transforming under a different group, $SU(2)_L$.  If the only fields in the theory are right-handed Weyl spinor fields, the Schwinger function cannot give an invariant inner product.
	\item After analytic continuation the self-adjoint factor  $E+\boldsymbol{\sigma}\cdot \mathbf p$ is neither self-adjoint nor skew-adjoint.  This makes it difficult to give $S_2$ an interpretation as inner product for a Euclidean field theory.
\end{itemize}

The first problem can be addressed by introducing fields of both chiralities, giving up on having a theory of only one chirality of Weyl spinors.  The adjointness problem however still remains. Schwinger and later authors have dealt with this problem by doubling the number of degrees of freedom.  Schwinger's argument was that this was necessary in order to have Euclidean transformation properties that did not distinguish a time direction.  The problem also appears when one tries to find a generalization of the time-reflection operator $\Theta$ that allows reconstruction of the Minkowski theory from the Euclidean theory.  The conventional wisdom has been to follow Osterwalder-Schrader, who deal with this by doubling the degrees of freedom, using a $\Theta$ which interchanges the two sorts of fields\cite{osterwalder-schrader}.  A fairly complete bibliography of attempts to deal with the Euclidean quantum spinor field is included at the end of this article.

\subsection{Physical states and $SO(4)$ symmetry breaking}

It appears to be a fundamental feature of Euclidean quantum field theory that, although Schwinger functions are $SO(4)$ invariant, recovering a connection to the physical theory in Minkowski space-time requires breaking  $SO(4)$ invariance by a choice of imaginary time direction.  In Minkowski space-time there is a Lorentz-invariant distinction between positive and negative energy, while in Euclidean space-time the corresponding distinction between positive and negative imaginary time is not $SO(4)$ invariant.  This breaking of $SO(4)$ symmetry is a sort of spontaneous symmetry breaking with not the lowest energy state, but the distinction between positive and negative energy necessary for quantization being responsible for the symmetry breaking.  

While the Euclidean Fock space has an $SO(4)$ action, states in it correspond not to physical states, but to paths in the space of physical states.   A choice of imaginary time direction is needed to get physical states, either by restriction to a constant imaginary time subspace or by restriction to a positive imaginary time subspace together with use of reflection in imaginary time.  The path integral formalism has the same feature: one can write Schwinger functions as an $SO(4)$ invariant path integral, but to get states one must choose a hypersurface and then define states using path integrals with fixed data on the hypersurface.

Needing to double spinor degrees of freedom and not being able to write down a free chiral spinor theory have always been disconcerting aspects of Euclidean quantum field theory.  An alternate interpretation of the problems with quantizing spinor fields in Euclidean space-time would be that they are a more severe version of the problem one already sees with relativistic scalars,  with the quantization of such theories requiring the introduction of a new degree of freedom that picks out an imaginary time direction.

\section{Hyperfunctions}
\label{s:hyperfunctions}

Wightman functions are conventionally described as tempered distributions on a Schwartz space of test functions.  Such distributions occur as boundary values of holomorphic functions, and one can instead work with hyperfunctions, which are spaces of such boundary values.  Like distributions, they can be thought of a elements of a dual space to a space of well-behaved test functions, which will be real analytic, not just infinitely differentiable. 
For an enlightening discussion of hyperfunctions in this context, a good source is chapter 9 of Roger Penrose's {\it  The Road to Reality} \cite{penrose}.

\subsection{Hyperfunctions on the circle}

In the case of the unit circle, one can generalize the notion of functions by considering boundary values of holomorphic functions on the open unit disk.   Taking the circle to be the equator of a Riemann sphere, a hyperfunction on the circle can be defined as a pair of functions, one holomorphic on the open upper hemisphere, the other holomorphic on the open lower hemisphere, with pairs equivalent when they differ by a globally holomorphic function.

Boundary values of functions holomorphic on the upper hemisphere correspond to Fourier series with Fourier coefficients satisfying $a_n=0$ for $n<0$,  those with $a_n=0$ for $n>0$ correspond to boundary values of functions holomorphic on the lower hemisphere.  The global holomorphic functions on the sphere are just the constants, those with only $a_0$ non-zero.  Hyperfunctions allow one to make sense of a very large class of Fourier series (those with coefficients growing at less than exponential rate as $n\rightarrow \pm \infty$) as linear functionals on real analytic test functions (whose coefficients $a_n$ fall off faster than $e^{-c|n|}$ for some $c>0$).

The discrete series representations of the non-compact Lie group $SL(2,\mathbf R)$ can naturally be constructed using such hyperfunctions on the circle.  The group $SL(2,\mathbf R)$ acts on the Riemann sphere, with orbits the upper hemisphere, the lower hemisphere, and the equator.  The discrete series representations are hyperfunctions on the equator, boundary values of holomorphic sections of a line bundle on either the upper or lower hemisphere.  For more about this, see section 10.1 of \cite{baston-eastwood}.  For a more general discussion of hyperfunctions on the circle and their relation to hyperfunctions on $\mathbf R$, see the previously mentioned chapter 9 of \cite{penrose}.

\subsection{Hyperfunctions on $\mathbf R$}

Solutions to wave equations are conventionally discussed using the theory of distributions, since even the simplest plane-wave solutions are delta-functions in energy-momentum space.  Distributions are generalizations of functions that can be defined as elements of the dual space (linear functionals) of some well-behaved set of functions, for instance  smooth functions of rapid decrease (Schwartz functions) for the case of tempered distributions.  The theory of hyperfunctions gives a further generalization, providing a dual of an even more restricted set of functions, analytic functions. Two references which contain extensive discussions of the theory of hyperfunctions with applications are \cite{imai} and \cite{graf}.   

To motivate the definition of a hyperfunction on $\mathbf R$, consider the boundary values of a holomorphic function  $\Phi_+$ on the open upper half plane.  These give a generalization of the usual notion of distribution, by considering the linear functional on analytic functions (satisfying an appropriate growth condition) on $\mathbf R$ 
$$f\rightarrow \lim_{\epsilon\rightarrow 0^+} \int_{-\infty}^\infty \Phi_+(t+i\epsilon)f(t)dt$$
Usual distributions are often written with a formal integral symbol denoting the linear functional.  In the case of hyperfunctions, this is no longer formal, but becomes (a limit of) a conventional integral of a holomorphic function in the complex plane, so contour deformation and residue theorem techniques can be applied to its evaluation.

It is sometimes more convenient to have a definition involving symmetrically the upper and lower complex half-planes.   The space $\mathcal B(\mathbf R)$ of hyperfunctions on $\mathbf R$ can be defined as equivalence classes of pairs of functions $(\Phi_+,\Phi_-)$, where $\Phi_+$ is a holomorphic function on the open upper half-plane, $\Phi_-$ is a holomorphic function on the open lower upper half-plane.   Pairs $(\Phi_{1,+},\Phi_{1,-})$ and $(\Phi_{2,+},\Phi_{2,-})$ are equivalent if
$$\Phi_{2,+}=\Phi_{1,+} + \psi,\ \ \Phi_{2,-}=\Phi_{1,-} +\psi$$ 
for some globally holomorphic function $\psi$.  We'll then write a hyperfunction as
$$\phi=[\Phi_+,\Phi_-]$$
The derivative $\phi^\prime$ of a hyperfunction $\phi$ is given by taking the complex derivatives of the pair of holomorphic functions  representing it
$$ \phi^\prime=[\Phi_+^\prime,\Phi_-^\prime]$$

As a linear functional on analytic functions, the hyperfunction $\phi$ is given by
$$f\rightarrow \oint _{-\infty}^{\infty} \phi(t)f(t)dt\equiv  \lim_{\epsilon\rightarrow 0^+}\int_{-\infty}^\infty(\Phi_+(t +i\epsilon)- \Phi_-(t -i\epsilon))f(t)dt$$
We'll use coordinates $t$ on $\mathbf R$, $z=t+i\tau$ on $\mathbf C$ since our interest will be in physical applications involving functions of time $t$, as well as their analytic continuations to imaginary time $\tau$.

One way to get hyperfunctions is by choosing a function $\Phi(z)$ on $\mathbf C$,  holomorphic away from the real axis $\mathbf R$, and taking
$$\phi=[\Phi_{|UHP}, \Phi_{|LHP}]$$
For example, consider the function 
$$\Phi=\frac{i}{2\pi}\frac{1}{z-\omega}$$
where $\omega\in \mathbf R$.  As a distribution, corresponding hyperfunction will be given by the limit
$$\phi(t)=\lim_{\epsilon\rightarrow 0^+}\frac{i}{2\pi} \left(\frac{1}{t+i\epsilon - \omega}-\frac{1}{t-i\epsilon - \omega}\right)=\lim_{\epsilon\rightarrow 0^+}\frac{1}{\pi}\frac{1}{(t-\omega)^2 + \epsilon^2}$$
The limit on the right-hand side is well-known as a way to describe the delta function distribution $\delta(t-\omega)$ as a limit of functions.  Using contour integration methods  one finds that the hyperfunction version
of the delta function  behaves as expected since
$$\oint _{-\infty}^{\infty}\frac{i}{2\pi}\frac{1}{t-\omega}f(t)dt=f(\omega)$$

One would like to define a Fourier transform for hyperfunctions, with the same sort of definition as an integral in the usual case, so
\begin{equation}
	\label{eq:fourier-time}
	\mathcal F(\phi)(E)=\widetilde{\phi}(E)=\frac{1}{\sqrt{2\pi}}\int _{-\infty}^{\infty}e^{iE t}\phi(t)dt
\end{equation}
with the inverse Fourier transform defined by
$$\mathcal F^{-1}(\widetilde{\phi})(t)=\phi(t)=\frac{1}{\sqrt{2\pi}}\int _{-\infty}^{\infty}e^{-iE t}\widetilde{\phi}(E)dE$$
The problem with this though is that the Fourier transform and its inverse don't take functions holomorphic on the upper or lower half plane to functions with the same property.

One can however define a Fourier transform for hyperfunctions (satisfying a growth condition, called \lq\lq Fourier hyperfunctions") by taking advantage of the fact that for a class of functions $f(E)$ supported on $E>0$  (respectively $E<0$) 
$$\frac{1}{\sqrt{2\pi}}\int_{-\infty}^\infty e^{-iEz}f(E)dE$$
is holomorphic in the lower half (respectively upper half) $z$ plane (since the exponential falls off there).    The decomposition of a hyperfunction $\phi(t)$ into limits of holomorphic functions $\Phi_+,\Phi_-$ on the upper and lower half planes corresponds to decomposition of $\widetilde{\phi}(E)$ into hyperfunctions $\widetilde{\phi}_-(E), \widetilde{\phi}_+(E)$  supported for negative and positive $E$ respectively.  This is similar to what happened for hyperfunctions on the circle, with $\Phi_+,\Phi_-$ analogous to functions holomorphic on the upper or lower hemispheres,  $\widetilde{\phi}_-(E), \widetilde{\phi}_+(E)$ analogous to the Fourier coefficients for positive or negative $n$.

For an example, consider the hyperfunction version of a delta function supported at $E=\omega, \omega >0$:
$$\widetilde {\phi}(E)=\widetilde{\phi}_+(E)=\frac{i}{2\pi}\frac{1}{E-\omega}\equiv \frac{i}{2\pi}\lim_{\epsilon\rightarrow 0^+}\left(\frac{1}{E+i\epsilon-\omega}-\frac{1}{E-i\epsilon-\omega}\right)$$
This has as inverse Fourier transform the hyperfunction
$$\phi(t)=\frac{1}{\sqrt{2\pi}}\int _{-\infty}^{\infty}\frac{i}{2\pi}\frac{1}{E-\omega}e^{-iE t}dE=\frac{1}{\sqrt{2\pi}}e^{-i\omega t}$$
which has a representation as
$$\phi(t)=[0,-\frac{1}{\sqrt{2\pi}}e^{-i\omega z}]$$
The Fourier transform of this will be
\begin{align*}
	\widetilde {\phi}(E)=&\frac{1}{\sqrt{2\pi}}\int _{-\infty}^{\infty}e^{iE t}\phi(t)dt\\
	=&\frac{1}{\sqrt{2\pi}}\int_{-\infty}^\infty e^{iE t}\frac{1}{\sqrt{2\pi}}e^{-i\omega t}dt
\end{align*}
but this needs to be interpreted as a sum of integrals for $t$ negative and $t$ positive
\begin{align*}
	=&\frac{1}{2\pi}\lim_{\epsilon\rightarrow 0^+}(\int_{-\infty}^0 e^{i(E-i\epsilon-\omega)t}dt + \int_0^\infty e^{i(E+i\epsilon-\omega)t}dt)\\
	=&\frac{i}{2\pi}\lim_{\epsilon\rightarrow 0^+}\left(\frac{1}{E+i\epsilon-\omega}-\frac{1}{E-i\epsilon-\omega}\right)
\end{align*}

An example that is relevant to the case of the harmonic oscillator is that of

$$\widetilde{\phi}(E)=\frac{1}{E^2-\omega^2}=\frac{1}{2\omega}\left(\frac{1}{E-\omega}-\frac{1}{E+\omega}\right)$$
where the first term is a hyperfunction with support only at $\omega >0$, the second only at $-\omega<0$.  The inverse Fourier transform is
$$\phi(t)=\frac{i\pi}{\omega}\frac{1}{\sqrt{2\pi}}(e^{i\omega t}-e^{-i\omega t})$$
where the first term should be interpreted as the equivalence class
$$\frac{i\pi}{\omega}\frac{1}{\sqrt{2\pi}}[e^{i\omega z},0]$$
and the second as the equivalence class
$$\frac{i\pi}{\omega}\frac{1}{\sqrt{2\pi}}[0,e^{-i\omega z}]$$

\printbibliography[notkeyword={euclidean},title={References}]
\printbibliography[keyword={euclidean},title={Euclidean Spinor References}]

@misc{adamo-ym,
	title={Twistor actions for gauge theory and gravity}, 
	author={Tim Adamo},
	year={2013},
	eprint={1308.2820},
	archivePrefix={arXiv},
	primaryClass={hep-th}
}

@INPROCEEDINGS{adamo,
       author = {{Adamo}, Timothy},
        title = "{Lectures on twistor theory}",
    booktitle = {Proceedings of the XIII Modave Summer School in Mathematical Physics. 10-16 September 2017 Modave},
         year = 2017,
        pages = {3},
archivePrefix = {arXiv},
       eprint = {1712.02196},
 primaryClass = {hep-th},
}

@article{alexander-marciano-smolin,
	title={Gravitational origin of the weak interaction’s chirality},
	volume={89},
	journal={Physical Review D},
	author={Alexander, Stephon and Marcianò, Antonino and Smolin, Lee},
	year={2014},
	pages={065017},
	archivePrefix = {arXiv},
	eprint = {1212.5246},
	primaryClass = {hep-th},
	
}

@book{ashtekar1,
author = {Ashtekar, Abhay},
title = {New Perspectives in Canonical Gravity},
year = 1988,
publisher = Bibliopolis
}

@book{ashtekar2,
author = {Ashtekar, Abhay},
title = {Lectures on Non-Perturbative Canonical Gravity},
publisher = {World Scientific},
year = {1991},
doi = {10.1142/1321},
}

@article{atiyah-hitchin-singer,
	AUTHOR = {Atiyah, Michael F. and Hitchin, Nigel J. and Singer, Isidore M.},
	TITLE = {Self-duality in four-dimensional Riemannian geometry},
	JOURNAL = {Proc. R. Soc. Lond. A},
	VOLUME = {362},
	YEAR = {1978},
	PAGES = {425-461},
}

@book{atiyah,
AUTHOR={Atiyah, Michael F.},
TITLE={Geometry of {Y}ang-{M}ills fields},
SERIES= {Lezioni Fermiane},
PUBLISHER ={Scuola Normale Superiore, Pisa},
YEAR ={1979},
}

@book{baston-eastwood,
AUTHOR={Baston, Robert J. and Eastwood, Michael G.},
TITLE={The {P}enrose Transform: Its Interaction With Representation Theory},
PUBLISHER ={Oxford University Press},
YEAR ={1989},
}

@article{boels-mason-skinner,
	title={Supersymmetric gauge theories in twistor space},
	volume={2007},
	journal={Journal of High Energy Physics},
	author={Boels, Rutger and Mason, Lionel and Skinner, David},
	year={2007},
	archivePrefix = {arXiv},
	eprint = {hep-th/0604040},
	primaryClass = {hep-th}
}

@article{borel-weil-bott,
    AUTHOR = {Bott, Raoul},
     TITLE = {Homogeneous Vector Bundles},
   JOURNAL = {Annals of Mathematics},
    VOLUME = {66},
      YEAR = {1957},
     PAGES = {203-248},
}

@incollection{bott,
  author       = {Bott, Raoul}, 
  title        = {The Index Theorem for Homogeneous Differential Operators},
  booktitle    = {Differential and Combianatorial Topology},
  publisher    = {Princeton University Press},
  year         = 1965,
  editor       = {Cairns, Stewart Scott},
  pages        = {167-186},
}

@article{chalmers-siegel,
	title={Self-dual sector of QCD amplitudes},
	volume={54},
	ISSN={1089-4918},
	journal={Physical Review D},
	author={Chalmers, G. and Siegel, W.},
	year={1996},
	pages={7628–7633},
	archivePrefix = {arXiv},
	eprint = {hep-th/9606061},
	primaryClass = {hep-th},
}

@misc{costello,
	title={Notes on supersymmetric and holomorphic field theories in dimensions 2 and 4}, 
	author={Kevin J. Costello},
	year={2013},
	eprint={1111.4234},
	archivePrefix={arXiv},
	primaryClass={math.QA}
}

@incollection{dubois-violette,
  author       = {Dubois-Violette, Michel}, 
  title        = {Structures Complexes Au-dessus des Vari\'et\'es, Applications},
  booktitle    = {Math\'ematique et Physique},
  publisher    = {Birkha\"user},
  year         = 1983,
  editor       = {Boutet de Montvel, L. and Douady, A. and Verdier, J.L.},
  volume       = 37,
  series       = {Progress in Mathematics},
  pages        = {1-42},
}

@article {epw,
    AUTHOR = {Eastwood, Michael G. and Penrose, Roger and Wells, Jr., Raymond O.},
     TITLE = {Cohomology and Massless Fields},
   JOURNAL = {Communications in Mathematical Physics},
    VOLUME = {78},
      YEAR = {1981},
     PAGES = {305-351},
}

@article {eastwood,
    AUTHOR = {Eastwood, Michael G.},
     TITLE = {The generalized {P}enrose-{W}ard transform},
   JOURNAL = {Proc. R. Soc. Lond. A},
    VOLUME = {370},
      YEAR = {1980},
     PAGES = {173-191},
}

@misc{elliott-yoo,
	title={Quantum Geometric Langlands Categories from N = 4 Super Yang-Mills Theory}, 
	author={Chris Elliott and Philsang Yoo},
	year={2020},
	eprint={2008.10988},
	archivePrefix={arXiv},
	primaryClass={math-ph}
}

@inproceedings{gibbons,
    author = "Gibbons, G.",
    title = "{Euclidean quantum gravity: The view from 2002}",
    booktitle = "{Workshop on Conference on the Future of Theoretical Physics and Cosmology in Honor of Steven Hawking's 60th Birthday}",
    pages = "351--372",
    year = "2002"
}

@book{glimm-jaffe,
	AUTHOR={Glimm, James and Jaffe, Arthur},
	TITLE={Quantum Physics: A Functional Integral Point of View},
	PUBLISHER ={Springer-Verlag},
	YEAR ={1981},
}

@ARTICLE{herfray,
       author = {{Herfray}, Yannick},
        title = "{Pure connection formulation, twistors, and the chase for a twistor action for general relativity}",
      journal = {Journal of Mathematical Physics},
         year = 2017,
       volume = {58},
       number = {11},
        pages = {112505},
archivePrefix = {arXiv},
       eprint = {1610.02343},
 primaryClass = {hep-th},
}

@article {hitchin,
    AUTHOR = {Hitchin, N. J.},
     TITLE = {Linear field equations on self-dual spaces},
   JOURNAL = {Math. Proc. Camb. Phil. Soc.},
    VOLUME = {97},
      YEAR = {1984},
     PAGES = {165-187},
}

@book{huang-pandzic,
AUTHOR={Huang, Jing-Song and Pand\u{z}i\'{c}, Pavle},
TITLE={Dirac Operators in Representation Theory},
PUBLISHER ={Birkh\"{a}user},
YEAR ={2006},
}

@article{kapustin-witten,
	author = "Kapustin, Anton and Witten, Edward",
	title = "{Electric-Magnetic Duality And The Geometric Langlands Program}",
	eprint = "hep-th/0604151",
	archivePrefix = "arXiv",
	journal = "Commun. Num. Theor. Phys.",
	volume = "1",
	pages = "1--236",
	year = "2007"
}

@article {klein-landau,
	AUTHOR = {Klein, Abel and Landau, Lawrence},
	TITLE = {From the Euclidean Group to the Poincar\'{e} Group via Osterwalder-Schrader Positivity},
	JOURNAL = {Commun. Math. Phys.},
	VOLUME = {87},
	YEAR = {1983},
	PAGES = {469-484},
}

@book{kobayashi-nomizu,
	AUTHOR={Kobayashi, Shoshichi and Nomizu, Katsumi},
	TITLE={Foundations of Differential Geometry: Volume I},
	PUBLISHER ={Interscience Publishers},
	YEAR ={1963},
}

@book{krasnov,
	AUTHOR={Krasnov, Kirill},
	TITLE={Formulations of General Relativity},
	PUBLISHER ={Cambridge University Press},
	YEAR ={2020},
}

@book {lawson-michelsohn,
    AUTHOR = {Lawson, Jr., H. Blaine and Michelsohn, Marie-Louise},
     TITLE = {Spin geometry},
    SERIES = {Princeton Mathematical Series},
    VOLUME = {38},
 PUBLISHER = {Princeton University Press},
      YEAR = {1989},
}

@book{mason-woodhouse,
AUTHOR={Mason, L. J. and Woodhouse, N.M.J.},
TITLE={Integrability, Self-duality, and Twistor Theory},
SERIES= {London Mathematical Society Mongraphs},
PUBLISHER ={Oxford University Press},
YEAR ={1996},
}

@article{nair,
	title = {A current algebra for some gauge theory amplitudes},
	journal = {Physics Letters B},
	volume = {214},
	pages = {215-218},
	year = {1988},
	author = {V.P. Nair}
}

@article{nesti-percacci,
	title={Gravi-weak unification},
	volume={41},
	journal={Journal of Physics A: Mathematical and Theoretical},
	publisher={IOP Publishing},
	author={Nesti, Fabrizio and Percacci, Roberto},
	year={2008},
	pages={075405},
	archivePrefix = {arXiv},
	eprint = "0706.3307",
	primaryClass = "hep-th",
}

@article {osterwalder-schrader,
    AUTHOR = {Osterwalder, Konrad and Schrader, Robert},
     TITLE = {Euclidean {F}ermi Fields and a {F}eynman-{Ka}c formula for {B}oson-{F}ermion models},
   JOURNAL = {Helvetica Physica Acta},
    VOLUME = {46},
      YEAR = {1973},
     PAGES = {277-302},
}

@article {penrose-twistors,
    AUTHOR = {Penrose, Roger},
     TITLE = {Twistor Algebra},
   JOURNAL = {Journal of Mathematical Physics},
    VOLUME = {8},
      YEAR = {1967},
     PAGES = {345-366},
}

@article {rawnsley,
    AUTHOR = {Rawnsley, John and Schmid, Wilfried and Wolf, Joseph A.},
     TITLE = {Singular Unitary Representations and Indefinite Harmonic Theory},
   JOURNAL = {Journal of Functional Analysis},
    VOLUME = {51},
      YEAR = {1983},
     PAGES = {1-114},
}

@incollection{salamon,
  author       = {Salamon, Simon}, 
  title        = {Topics in Four-dimensional {R}iemannian geometry},
  booktitle    = {Geometry Seminar Luigi Bianchi},
  publisher    = {Springer},
  year         = 1983,
  editor       = {Vesentini, E.},
  volume       = 1022,
  series       = {Lecture Notes in Mathematics},
  pages        = {33-124},
}

@book {ward-wells,
    AUTHOR = {Ward, R. S. and Wells, Jr., Raymond O.},
     TITLE = {Twistor geometry and field theory},
 PUBLISHER = {Cambridge University Press},
      YEAR = {1990},
}

@article {wells-hyperfunctions,
    AUTHOR = {Wells, Jr., R. O.},
     TITLE = {Hyperfunction Solutions of the Zero-Rest-Mass Field Equations},
   JOURNAL = {Communications in Mathematical Physics},
    VOLUME = {78},
      YEAR = {1981},
     PAGES = {567-600},
}

@article{witten-tqft,
       author = {{Witten}, Edward},
        title = "{Topological quantum field theory}",
      journal = {Communications in Mathematical Physics},
         year = 1988,
       volume = {117},
       number = {3},
        pages = {353-386},
}

@ARTICLE{witten-twistor,
       author = {{Witten}, Edward},
        title = "{Perturbative Gauge Theory as a String Theory in Twistor Space}",
      journal = {Communications in Mathematical Physics},
         year = 2004,
       volume = {252},
       number = {1-3},
        pages = {189-258},
archivePrefix = {arXiv},
       eprint = {hep-th/0312171},
 primaryClass = {hep-th},
}

@article {woit-ssqm,
    AUTHOR = {Woit, Peter},
     TITLE = {Supersymmetric quantum mechanics, spinors and the Standard Model},
   JOURNAL = {Nuclear Physics},
    VOLUME = {B303},
      YEAR = {1988},
     PAGES = {329-342},
}

@misc{woit-gaugesymmetry,
	title={Quantum Field Theory and Representation Theory: A Sketch}, 
	author={Peter Woit},
	year={2002},
	archivePrefix = {arXiv},
	eprint = {hep-th/0206135},
	primaryClass = {hep-th}
}

@book {woit-qmbook,
    AUTHOR = {Woit, Peter},
     TITLE = {Quantum theory, groups and representations: An introduction},
 PUBLISHER = {Springer},
      YEAR = {2017},
}

@article {woodhouse,
    AUTHOR = {Woodhouse, N.M.J.},
     TITLE = {Real Methods in Twistor Theory},
   JOURNAL = {Classical and Quantum Gravity},
    VOLUME = {2},
      YEAR = {1985},
     PAGES = {257-291},
}

@book {graf,
    AUTHOR = {Graf, Urs},
     TITLE = {Introduction to hyperfunctions and their integral transforms},
 PUBLISHER = {Birkh{\"a}user},
      YEAR = {2010},
}

@book {imai,
    AUTHOR = {Imai, Isao},
     TITLE = {Applied hyperfunction theory},
    SERIES = {Mathematics and its applications},
    VOLUME = {8},
 PUBLISHER = {Kluwer Academic Publishers},
      YEAR = {1992},
}

@book {penrose,
  AUTHOR = {Penrose, Roger},
  TITLE = {The Road to Reality},
  YEAR = {2004},
  PUBLISHER = {Jonathan Cape},
}

@article {schwinger1958-1,
	author = {Schwinger, Julian},
	title = {On the Euclidean structure of relativistic field theory},
	volume = {44},
	number = {9},
	pages = {956--965},
	year = {1958},
	journal = {Proceedings of the National Academy of Sciences}
}

@inproceedings{schwinger1958-2,
	author = "Schwinger, Julian",
	title = "{Four-dimensional Euclidean formulation of quantum field theory}",
	booktitle = "{8th International Annual Conference on High Energy Physics}",
	pages = "134--140",
	year = "1958"
}

@article{PhysRev.115.721,
	title = {Euclidean Quantum Electrodynamics},
	author = {Schwinger, Julian},
	journal = {Phys. Rev.},
	volume = {115},
	issue = {3},
	pages = {721--731},
	year = {1959}
}

@article{10.1143/PTP.21.241,
	author = {Nakano, Tadao},
	title = "{Quantum Field Theory in Terms of Euclidean Parameters}",
	journal = {Progress of Theoretical Physics},
	volume = {21},
	number = {2},
	pages = {241-259},
	year = {1959}
}

@inproceedings{Osterwalder:1973zw,
	author = "Osterwalder, K.",
	title = "{Euclidean fermi fields}",
	booktitle = "{International School of Mathematical Physics, Ettore Majorana: 1st course: Constructive Quantum Field Theory}",
	pages = "326--331",
	year = "1973"
}

@article{osterwalder1973-2,
	author = "Osterwalder, K. and Schrader, R.",
	title = "{Euclidean fermi fields and a Feynman-Kac formula for boson-fermion models}",
	journal = "Helv. Phys. Acta",
	volume = "46",
	pages = "277--302",
	year = "1973"
}

@article{williams1974,
	author = "Williams, David N.",
	fjournal = "Communications in Mathematical Physics",
	journal = "Comm. Math. Phys.",
	number = "1",
	pages = "65--80",
	publisher = "Springer",
	title = "Euclidean Fermi fields with a Hermitean Feynman-Kac-Nelson formula, I",
	volume = "38",
	year = "1974"
}

@unpublished{williams1974a,
	author = "Williams, David N.",
	title = "Euclidean Fermi fields with a Hermitean Feynman-Kac-Nelson formula, II ",
	url = "http://www-personal.umich.edu/~williams/papers/eufermi.pdf",
	year = "1974"
}

@inproceedings {nelson1974,
	AUTHOR = {Nelson, Edward},
	TITLE = {Markov fields},
	BOOKTITLE = {Proceedings of the {I}nternational {C}ongress of
	{M}athematicians ({V}ancouver, {B}. {C}., 1974), {V}ol. 2},
	PAGES = {395--398},
	YEAR = {1975}
}

@article{Frohlich:1974zs,
	author = "Frohlich, Jurg and Osterwalder, Konrad",
	title = "{Is There a Euclidean Field Theory for Fermions?}",
	journal = "Helv. Phys. Acta",
	volume = "47",
	pages = "781",
	year = "1975"
}

@article{GROSS1977162,
	title = "On the formula of Mathews and Salam",
	journal = "Journal of Functional Analysis",
	volume = "25",
	number = "2",
	pages = "162 - 209",
	year = "1977",
	author = "Leonard Gross"
}

@article{NICOLAI1978294,
	title = "A possible constructive approach to $(\text{super}-\phi^3)_4$: (I). Euclidean formulation of the model",
	journal = "Nuclear Physics B",
	volume = "140",
	number = "2",
	pages = "294 - 300",
	year = "1978",
	issn = "0550-3213",
	author = "H. Nicolai"
}

@article{doi:10.1063/1.524533,
	author = {de Falco,Diego  and Guerra,Francesco },
	title = {On the local structure of the Euclidean Dirac field},
	journal = {Journal of Mathematical Physics},
	volume = {21},
	number = {5},
	pages = {1111-1114},
	year = {1980}
}

@article{PALMER1980287,
	title = "Euclidean Fermi fields",
	journal = "Journal of Functional Analysis",
	volume = "36",
	number = "3",
	pages = "287 - 312",
	year = "1980",
	author = "John Palmer"
}

@book{ramond,
	AUTHOR={Ramond, Pierre},
	TITLE={Field Theory: A Modern Primer},
	PUBLISHER ={Benjamin/Cummings},
	YEAR ={1981},
}

@article{10.1143/PTP.66.1061,
	author = {Nagamachi, Shigeaki and Mugibayashi, Nobumichi},
	title = "{Covariance of Euclidean Fermi Fields}",
	journal = {Progress of Theoretical Physics},
	volume = {66},
	number = {3},
	pages = {1061-1077},
	year = {1981}
}

@article{ek1981-1,
	author = {Ek, Bengt},
	title = {Parity Conservation and the Euclidean Field Formulation of Relativistic Quantum Theoriesj},
	journal = {Letters in Mathematical Physics},
	volume = {5},
	year = {1981},
	pages = {149-154}
}

@article{ek1981-2,
	author = {Ek, Bengt},
	title ={Euclidean Quantum Fields with Spin on an Indefinite Inner Product State Space},
	journal = {Publ. RIMS, Kyoto Univ.},
	volume = {18},
	year = {1982},
	pages = {251-274}
}

@article{frohlich-osterwalder-seiler,
	author = {J. Frohlich and K. Osterwalder and E. Seiler},
	journal = {Annals of Mathematics},
	number = {3},
	pages = {461--489},
	publisher = {Annals of Mathematics},
	title = {On Virtual Representations of Symmetric Spaces and Their Analytic Continuation},
	volume = {118},
	year = {1983}
}

@article{doi:10.1002/prop.2190350503,
	author = {Kupsch, J.},
	title = {Measures for Fermionic Integration},
	journal = {Fortschritte der Physik/Progress of Physics},
	volume = {35},
	number = {5},
	pages = {415-436},
	year = {1987}
}

@article{AIHPA_1989__50_2_143_0,
	author = {Kupsch, J.},
	title = {Functional integration for Euclidean Dirac fields},
	journal = {Annales de l'I.H.P. Physique th\'eorique},
	publisher = {Gauthier-Villars},
	volume = {50},
	number = {2},
	year = {1989},
	pages = {143-160}
}

@article{doi:10.1002/prop.2190380103,
	author = {Kupsch, J. and Thacker, W. D.},
	title = {Euclidean Majorana and Weyl Spinors},
	journal = {Fortschritte der Physik/Progress of Physics},
	volume = {38},
	number = {1},
	pages = {35-62},
	year = {1990}
}

@article{PhysRevLett.65.1983,
	title = {Euclidean continuation of the Dirac fermion},
	author = {Mehta, Mayank R.},
	journal = {Phys. Rev. Lett.},
	volume = {65},
	issue = {16},
	pages = {1983--1986},
	year = {1990},
	month = {Oct},
}

@ARTICLE{1993JMP....34.2691B,
	author = {{Borthwick}, David},
	title = "{Euclidean {M}ajorana fermions, fermionic integrals, and relative Pfaffians}",
	journal = {Journal of Mathematical Physics},
	year = 1993,
	month = jul,
	volume = {34},
	number = {7},
	pages = {2691-2712},
}

@article{van_Nieuwenhuizen_1996,
	title={On Euclidean spinors and Wick rotations},
	volume={389},
	number={1},
	journal={Physics Letters B},
	publisher={Elsevier BV},
	author={van Nieuwenhuizen, Peter and Waldron, Andrew},
	year={1996},
	month={Dec},
	pages={29–36}
}

@article{van_Nieuwenhuizen_1997,
	title={A continuous Wick rotation for spinor fields and supersymmetry in Euclidean space},
	journal={Gauge Theories, Applied Supersymmetry and Quantum Gravity II},
	publisher={PUBLISHED BY IMPERIAL COLLEGE PRESS AND DISTRIBUTED BY WORLD SCIENTIFIC PUBLISHING CO.},
	author={van Nieuwenhuizen, P. and Waldron, A.},
	year={1997},
	eprint ="hep-th/9608174",
	month={Jun}
}

@article{Waldron_1998,
	title={A Wick rotation for spinor fields: the canonical approach},
	volume={433},
	number={3-4},
	journal={Physics Letters B},
	publisher={Elsevier BV},
	author={Waldron, Andrew},
	year={1998},
	month={Aug},
	eprint="hep-th/9702057",
	pages={369–376}
}

@inproceedings{mountain,
	author = {Mountain, Arthur},
	title = {Wick rotation and supersymmetry},
	booktitle = {Quantum aspects of gauge theories, supersymmetry and unification},
	eventdate = {1999-09-01/1999-09-07},
	location = {Paris},
	year = {2000},
	doi = {10.22323/1.004.0036}
}

@article{Wessels_2004,
	title={Stability of Covariant Relativistic Quantum Theory},
	volume={35},
	number={1-2},
	journal={Few-Body-Systems},
	publisher={Springer Science and Business Media LLC},
	author={Wessels, V. and Polyzou, W. N.},
	year={2004},
	month={Sep}
}

@article{jaffe2008,
	title={Reflection positivity and monotonicity},
	volume={49},
	number={5},
	journal={Journal of Mathematical Physics},
	publisher={AIP Publishing},
	author={Jaffe, Arthur and Ritter, Gordon},
	year={2008},
	month={May},
	archivePrefix = "arXiv",
	eprint = "0705.0712",
	primaryClass = "math-ph",
	pages={052301}
}

@article{WETTERICH2011174,
	title = "Spinors in euclidean field theory, complex structures and discrete symmetries",
	journal = "Nuclear Physics B",
	volume = "852",
	number = "1",
	pages = "174 - 234",
	year = "2011",
	author = "C. Wetterich",
	archivePrefix = "arXiv",
	eprint = "1002.3556",
	primaryClass = "hep-th"
}

\end{document}